\documentclass[aps,prb,twocolumn,floatfix,longbibliography,nofootinbib]{revtex4-2}
\usepackage[T1]{fontenc}
\usepackage{amsmath,amssymb,bm}
\usepackage{graphicx}
\usepackage{booktabs}
\usepackage{dcolumn}
\usepackage[dvipsnames]{xcolor}
\usepackage[colorlinks=true,linkcolor=blue,citecolor=blue,urlcolor=blue]{hyperref}

\newcommand{\qv}{\mathbf{q}}
\newcommand{\Qv}{\mathbf{Q}}
\newcommand{\kv}{\mathbf{k}}
\newcommand{\Sxx}{S^{xx}}
\newcommand{\Szz}{S^{zz}}
\newcommand{\SpmPRL}{S^{+-}}
\newcommand{\Gxx}{G^{xx}}
\newcommand{\Gzz}{G^{zz}}
\newcommand{\rhos}{\rho_s}
\newcommand{\chizz}{\chi_{zz}}
\newcommand{\DIS}{\Delta_{\rm IS2}}
\newcommand{\Drot}{\Delta_{\rm rot}}

\begin{document}

\title{Zone-folded Goldstone mode of the field-induced spin supersolid in the square-lattice
       \texorpdfstring{$S=1$}{S=1} antiferromagnet}

\author{Sibin Yang}
\email{sibin.yang@ntu.edu.sg}
\affiliation{School of Physical and Mathematical Sciences, Nanyang Technological University, 21 Nanyang Link, Singapore 637371}

\author{Pinaki Sengupta}
\email{psengupta@ntu.edu.sg}
\affiliation{School of Physical and Mathematical Sciences, Nanyang Technological University, 21 Nanyang Link, Singapore 637371}

\date{\today}

\begin{abstract}
The field-induced spin supersolid of the square-lattice $S=1$ antiferromagnet with single-ion and exchange anisotropy has so far been characterized only by static observables, among them a two-peak transverse structure factor. We compute its dynamic spin structure factors, and those of the neighboring phases, by stochastic analytic continuation of stochastic-series-expansion quantum Monte Carlo data on a $16\times16$ lattice at inverse temperature $\beta=32$, at one field in each of the supersolid, the $m_z=1/2$ Ising plateau and the canted $XY$ phase. The supersolid has a single Goldstone mode, which the checkerboard folds onto both the ordering wave vector $\Qv=(\pi,\pi)$ and $\qv=(0,0)$, so that it appears in both spin channels at both points with one velocity $v=1.36\pm0.01$. The canted phase shows half of this pattern and the plateau none, the latter carrying instead a gapped magnon band and a two-magnon longitudinal continuum. Exactly at $\Qv$ and $(0,0)$ the finite-lattice spectra are governed by a rotor tower of the condensate in the transverse channel and an Ising doublet of the checkerboard in the longitudinal one, both present at $\Qv$ in the supersolid. Three parameter-free comparisons with the stiffness and the uniform susceptibility of the same runs, for the velocity, the sound-mode weight and the thermal decay of the rotor tower, tie the continued spectra to raw Monte Carlo observables. The folded Goldstone mode is the dynamical fingerprint of the supersolid.
\end{abstract}

\maketitle

\section{Introduction}
\label{sec:intro}

A supersolid breaks translational and $U(1)$ phase symmetry at the same time, so that a crystal carries a dissipationless superflow~\cite{Andreev1969,Chester1970,Leggett1970,Boninsegni2012}. After the inconclusive search in solid helium~\cite{Kim2004,Boninsegni2012}, the phase has been realized in cavity-coupled and dipolar quantum gases~\cite{Leonard2017,Tanzi2019,Guo2019,Natale2019}, and lattice supersolids of hard-core bosons were established numerically on frustrated lattices~\cite{Wessel2005,Heidarian2005,Melko2005,Boninsegni2005}. Quantum magnets offer a further route: a spin system in a field maps onto lattice bosons whose chemical potential is the field~\cite{Matsubara1956,Giamarchi2008,Zapf2014}, and a ``spin supersolid'' has coexisting diagonal (Ising) and off-diagonal ($XY$) long-range order. Hard-core ($S=1/2$) spin supersolids need frustration, either geometric~\cite{Yamamoto2014} or from competing further-neighbor interactions~\cite{Batrouni2000}, typically together with an easy-axis exchange anisotropy, and triangular-lattice cobaltates such as Na$_2$BaCo(PO$_4$)$_2$ and K$_2$Co(SeO$_3$)$_2$ are now studied as candidates~\cite{Gao2022,Xiang2024,Sheng2025,Zhu2024}.

Sengupta and Batista showed that frustration is not needed for $S=1$ spins~\cite{Sengupta2007} (hereafter the Letter), because an $S=1$ site can hold a second boson, which a hard-core boson site cannot. In the square-lattice $S=1$ antiferromagnet with a planar single-ion anisotropy $D$ and an easy-axis exchange anisotropy $\Delta$, a field drives the system from a N\'eel solid (IS1) through a supersolid (SS) into a second Ising plateau (IS2) at magnetization $m_z=1/2$ and then, by a first-order transition, into a canted $XY$ antiferromagnet. The same mechanism operates in the low-energy sector of frustrated $S=1$ dimer systems~\cite{Sengupta2007} and in anisotropic $S=1$ chains~\cite{Sengupta2007b,Peters2009,Rossini2011}; spin-$1/2$ dimer bilayers with a strongly Ising-like interdimer exchange host a supersolid as well~\cite{Ng2006,Laflorencie2007}.

The evidence for the $S=1$ supersolid is entirely static: magnetization plateaus, the staggered longitudinal structure factor, the condensate fraction, the spin stiffness, and the two-peak momentum distribution of the transverse structure factor~\cite{Sengupta2007}. To our knowledge its excitation spectrum has not been computed. The closest calculation, generalized spin-wave theory of a frustrated spin-$1/2$ dimer model whose low-energy sector nearly coincides with the $S=1$ model, finds a single Goldstone mode that is visible in the longitudinal response near both $(0,0)$ and $(\pi,\pi)$ and has the hydrodynamic velocity~\cite{Murakami2013}. What dynamics adds is a set of quantitative constraints on the coexisting orders, and symmetry fixes what they are. The supersolid breaks one continuous symmetry, the $U(1)$ rotation about the field, and one discrete symmetry, the $Z_2$ choice of sublattice. Only the former produces a Goldstone mode~\cite{Goldstone1961,Watanabe2012}; the discrete breaking costs a gapped domain wall and leaves on a finite lattice only an exponentially split Ising doublet. Unlike a continuum supersolid, whose continuously broken translations add phonon branches~\cite{Leonard2017,Tanzi2019,Guo2019,Natale2019}, the lattice supersolid therefore has exactly one gapless mode, and with a single broken generator that mode is of type A and disperses linearly even though the magnetization is finite~\cite{Watanabe2012}. The checkerboard doubles the unit cell, so this one mode should appear both at the ordering wave vector $\Qv=(\pi,\pi)$ and at $\qv=0$, with a single velocity that hydrodynamics fixes by the ratio of the stiffness to the uniform susceptibility, $v^2=\rhos/\chizz$~\cite{Halperin1969}. Neither the plateau (no $U(1)$ order) nor the canted $XY$ phase (no $Z_2$ order) can produce this pattern.

We compute the transverse and longitudinal dynamic structure factors of the model of Fig.~1 of the Letter on a $16\times16$ lattice, using stochastic series expansion (SSE) quantum Monte Carlo (QMC)~\cite{Sandvik1991,Sandvik1999,Syljuasen2002} and stochastic analytic continuation (SAC)~\cite{Sandvik1998,Beach2004,Sandvik2016,Shao2017,Shao2023}, at one field in each of the three ordered phases above the N\'eel solid. The supersolid shows the predicted doubled Goldstone mode in both channels with a single velocity, the $XY$ phase half of the pattern and the plateau none (Sec.~\ref{sec:doubling}), and three parameter-free comparisons with the static observables of the same runs tie the spectra to raw Monte Carlo data (Sec.~\ref{sec:checks}).

\section{Model and phases}
\label{sec:model}

\subsection{Hamiltonian and parameters}
\label{sec:params}

We study
\begin{equation}
  \mathcal H = J\sum_{\langle ij\rangle}\bigl(S^x_iS^x_j+S^y_iS^y_j+\Delta\,S^z_iS^z_j\bigr)
  + \sum_i\bigl[D\,(S^z_i)^2 - B\,S^z_i\bigr],
  \label{eq:H}
\end{equation}
with $S=1$ on the square lattice, antiferromagnetic transverse ($J$) and longitudinal ($J_z\equiv\Delta J$) couplings, a planar single-ion anisotropy $D>0$, and a field $B$. We set $J=1$, so that $J_z$ and $\Delta$ are the same number; we write $\Delta$ when quoting the Letter and $J_z$ when the coupling enters an energy.

We note that the QMC simulations in the Letter missed the Clebsch-Gordan coefficient in the transverse component of the exchange term. This results in an effective transverse coupling of $J/2$, which is equivalent, up to the overall energy unit, to doubling $D$, $J_z$ and $B$ at fixed $J$. In other words, the correct values of the parameters that yield the phase diagram in Fig.~1 of the Letter are twice the values quoted there. Accordingly, we work at $D=3.0$, $J_z=3.6$, the correct parameters of Fig.~1 of the Letter in the units of Eq.~\eqref{eq:H}.
This sets the lower edge of the supersolid at $B\simeq10.4$; our scans bracket that edge between $10.0$ and $10.5$ and give the two plateau edges $13.0$ and $14.8$. Appendix~\ref{app:scan} validates the simulations against exact diagonalization and against the phase sequence, the phase boundaries and the stiffness of the Letter. All results below are for $B=11.5$, $14$ and $17$, one field in each of the supersolid, the IS2 plateau and the canted $XY$ phase, with fields in the units of Eq.~\eqref{eq:H}; all energies are in units of $J$.

\subsection{Phases}
\label{sec:phases}

\begin{figure}[t]
\centering
\includegraphics[width=\columnwidth]{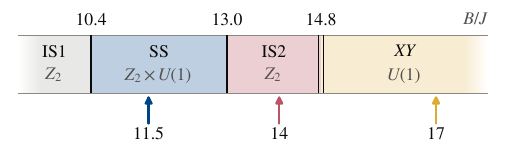}
\caption{Phases of Eq.~\eqref{eq:H} at $D=3.0$, $J_z=3.6$ along the field axis, with the symmetries each breaks: the $Z_2$ sublattice symmetry of the checkerboard and the $U(1)$ rotation about the field. Single lines mark the continuous transitions (the IS1--SS one according to the Letter) and the double line the first-order one, at the fields quoted in Sec.~\ref{sec:params} (scans in Appendix~\ref{app:scan}). Arrows: the three fields studied, colored as in Figs.~\ref{fig:summary}--\ref{fig:hydro}.}
\label{fig:phases}
\end{figure}

At $D=3.0$, $J_z=3.6$ the ground state passes through four phases as the field increases (Fig.~\ref{fig:phases}). Below $B\simeq10.4$ it is the gapped N\'eel Ising solid (IS1), with spins mostly in $S^z=\pm1$ on the two sublattices, $m_z=0$ and $\rhos=0$. Between $10.4$ and $13.0$ a continuous Bose-Einstein condensation~\cite{Sengupta2007} puts a finite density of $S^z=0$ sites preferentially on the $S^z=-1$ sublattice, and diagonal order [$\Szz(\Qv)/N$ finite, $N$ being the number of sites] coexists with off-diagonal order [$\Sxx(\Qv)/N$ finite] and a finite stiffness $\rhos$: this is the supersolid, whose transverse structure factor has Bragg-like peaks at both $\qv=0$ and $\Qv$ (the two-peak diagnostic of the Letter). Between $13.0$ and $14.8$ all $S^z=-1$ sites have become $S^z=0$, giving the second Ising solid (IS2), a plateau with $m_z=1/2$ exactly and vanishing stiffness. Above $14.8$ a first-order transition ($m_z:0.5\to0.59$) leads into a uniform canted $XY$ antiferromagnet.

In the boson language, $S^z_i=n_i-1$ maps Eq.~\eqref{eq:H} onto semi-hard-core bosons ($n_i\le2$) with nearest-neighbor repulsion $V=J_z$, hopping of order $J$, an on-site repulsion $U=2D$ in the normalization $\tfrac U2\sum_in_i(n_i-1)$ of Eq.~(5) of the Letter, and chemical potential $\mu=D+B$, with the repulsion written as $V(n_i-1)(n_j-1)$. The field controls the cost of putting a boson on the depleted sublattice, $E_0=zV-\mu=4J_z-D-B$ ($z=4$ neighbors), which vanishes at $B=zJ_z-D=11.4$, the field around which the supersolid window lies. At $B=11.5$, $E_0=-0.1$: turning a depleted site from $S^z=-1$ to $0$ gains $0.1$, and the reverse flip costs $0.1$. (Such local costs, used again below, are the diagonal energies of one changed site in the classical checkerboard.) The condition $|zV-U|\lesssim t$ that the Letter discusses for its dimer model is a different one: here $zV-U=8.4$, and it is the field that tunes the system into the window. The ingredient the two share, and the reason $S=1/2$ cannot do this, is the $n=2$ state that the added boson needs~\cite{Batrouni2000,Sengupta2005}.

$E_0$ alone would shrink the window to the single field $B=11.4$. It has a finite width because a flipped site can move, which lowers its energy. Two sites of the depleted sublattice are never neighbors, so a move takes two transverse steps through a common neighbor $a$. For an $S^z=0$ site $b$ in the IS1 checkerboard and a second depleted site $b'$, $(S^z_b,S^z_a,S^z_{b'})$ goes $(0,+1,-1)\to(0,0,0)\to(-1,+1,0)$. Each step has amplitude $J$ for $S=1$, and the intermediate state, in which $a$ has left $+1$, costs $\delta E=6J_z-2D=15.6$, so the flipped site hops with $t_2=J^2/\delta E=0.064$. For an $S^z=-1$ site on the plateau fewer bonds are broken: $\delta E=4J_z-2D=8.4$ and $t_2=0.12$. On its sublattice the site hops by $2t_2$ to its four diagonal neighbors (two paths each) and by $t_2$ to the four sites two steps away along an axis (one path), which gives a band of width $16t_2$ ($1.0$ and $1.9$) whose bottom lies $12t_2$ below the local cost. Flipped sites therefore enter once $E_0-12t_2$ ($S^z=0$ sites in IS1) or $-E_0-12t_2$ ($S^z=-1$ sites on the plateau) falls below zero, at $B\simeq10.6$ and $12.8$. The same virtual steps also shift the local costs, by $-0.12$ and $+0.17$, which moves these fields to $10.5$ and $12.7$; interactions between flipped sites are neglected. The edges are $10.4$ (the value of the Letter in our units, bracketed between $10.0$ and $10.5$ by our scan) and the measured $13.0$. The same band, $1$--$2$ wide, and not $|E_0|$, sets the energy scale of the Goldstone branch (Sec.~\ref{sec:SS}).

Throughout, the transverse structure factor is the symmetric combination $\Sxx=\langle S^xS^x\rangle=\tfrac14\langle S^+S^-+S^-S^+\rangle$. Although Eq.~(3) of the Letter is written as $\langle S^+_jS^-_k\rangle$, the quantity computed there, as in our code, is this symmetric combination; we write $\SpmPRL$ only when quoting the Letter.

\section{Methods}
\label{sec:method}

\subsection{Simulations and correlators}
\label{sec:sse}

\begin{table}[!tb]
\caption{Equal-time structure factors at the three fields ($L=16$, $\beta=32$) and the order parameters they encode [$m_z$ from $\Szz(0,0)=Nm_z^2$; solid order $\Szz(\Qv)/N$; condensate $\Sxx(\Qv)/N$; folded condensate $\Sxx(0,0)/N$], the stiffness and uniform susceptibility from the same runs, and the two quantities they predict, the hydrodynamic velocity and the rotor gap (Secs.~\ref{sec:checks} and \ref{sec:sectors}). $\rhos$ and $\chizz$ are per site and carry statistical errors below $10^{-4}$; on the plateau $\chizz<10^{-4}$ and $\rhos$ is at its finite-size background. Rows marked Ref.~\cite{Sengupta2007} are read from Fig.~1 of the Letter at the same field ($B/2$ on its axis), or, for panels (c) and (d), which are plotted against $m_z$, at the same magnetization, with a reading uncertainty of about $\pm0.03$ on the axes of that figure (for $\rhos$ the range read is given); its $\rhos$, in the energy unit of that figure (Sec.~\ref{sec:params}), is half of ours. Ratios are formed from the unrounded values; a dash means the quantity is not defined in that phase.}
\label{tab:static}
\begin{ruledtabular}
\begin{tabular}{lccc}
 & $B=11.5$ & $B=14$ & $B=17$\\
 & SS & IS2 & $XY$\\
\hline
$m_z$ & 0.360 & 0.500 & 0.706\\
\quad Ref.~\cite{Sengupta2007}, Fig.~1(a) & $\approx0.35$ & 0.5 & $\approx0.70$\\
$\Szz(\Qv)/N$ & 0.323 & 0.200 & 0.002\\
\quad Ref.~\cite{Sengupta2007}, Fig.~1(c) & $\approx0.35$ & $\approx0.2$ & 0\\
$\Sxx(\Qv)/N$ & 0.123 & 0.011 & 0.148\\
\quad Ref.~\cite{Sengupta2007}, Fig.~1(b) & $\approx0.13$ & $\to0$ & $\approx0.15$\\
$\Sxx(0,0)/N$ & 0.0257 & 0.0027 & 0.0015\\
$\Sxx(0,0)/\Sxx(\Qv)$ & 0.21 & 0.24 & 0.01\\
\hline
$\rhos$ & 0.2508 & 0.0006 & 0.4372\\
\quad Ref.~\cite{Sengupta2007}, Fig.~1(d) & $0.10$--$0.13$ & 0 & $0.2$--$0.3$\\
$\chizz$ & 0.1355 & $<10^{-4}$ & 0.0510\\
$\sqrt{\rhos/\chizz}$ & 1.36 & -- & 2.93\\
$\Drot=1/(2\chizz N)$ & 0.0144 & -- & 0.0383\\
\end{tabular}
\end{ruledtabular}
\end{table}

We simulate Eq.~\eqref{eq:H} with the SSE method~\cite{Sandvik1991,Sandvik1999,Syljuasen2002} on periodic $L\times L$ lattices with $L=16$ ($N=256$) at inverse temperature $\beta=32$. Every excitation we resolve has $\beta\omega\ge16$, so the spectra are ground-state properties up to corrections of order $e^{-\beta\omega}$, except for the symmetry-restoring sectors at the two special wave vectors, whose thermal occupation is treated explicitly (Secs.~\ref{sec:sectors} and \ref{sec:checks}), and for a shift of the transverse Goldstone peaks by the level spacing of those sectors (Sec.~\ref{sec:checks}). The imaginary-time correlators
\begin{equation}
  \Gxx(\qv,\tau)=\tfrac14\bigl\langle S^+_\qv(\tau)S^-_{-\qv}(0)+S^-_\qv(\tau)S^+_{-\qv}(0)\bigr\rangle
\end{equation}
and $\Gzz(\qv,\tau)=\langle S^z_\qv(\tau)S^z_{-\qv}(0)\rangle$ were measured for $0\le\tau\le\beta/2$ on a $9\times9$ momentum grid ($q_{x,y}=0,\pi/8,\dots,\pi$) in separate runs. The transverse correlator uses the time-sliced worm estimator of the off-diagonal Green's function of Ref.~\cite{Flynn2026}, in the boundary-uniform insertion variant of its Eq.~(50), which builds on the directed-loop measurement of Ref.~\cite{Dorneich2001}; the worm proposes $S^+$ and $S^-$ heads with equal probability and the two directed channels are not stored separately, so only the symmetric combination is available. Each measurement bin is $10^5$ Monte Carlo steps with a measurement at every step, the runs being continued from stored equilibrated configurations; there are $4096$ bins per channel at $B=11.5$ and $14$, and $4086$ ($\Gxx$) and $3875$ ($\Gzz$) at $B=17$, rebinned by two before continuation. The imaginary-time grid follows the slicing of the estimator, $\delta\tau=0.1$: uniform up to $\tau=5$ and then widening quadratically, it gives at most 61 points between $\tau=0.1$ and $15.6$. The same runs give the static observables of Table~\ref{tab:static}: $m_z$, the uniform susceptibility $\chizz=\beta[\langle(S^z_{\rm tot})^2\rangle-\langle S^z_{\rm tot}\rangle^2]/N$, the stiffness $\rhos=\langle W_x^2+W_y^2\rangle/(2\beta)$ from the winding numbers, and $\Szz(\Qv)$, pooled over the runs of both channels; $\chizz$ and $\rhos$ are per site.

\subsection{Analytic continuation}
\label{sec:sac}

The spectra were obtained with SAC~\cite{Sandvik1998,Beach2004,Sandvik2016,Shao2017,Shao2023,YangThesis}, a stochastic generalization of the maximum-entropy method~\cite{Jarrell1996,Beach2004}, with $2000$ $\delta$ functions free in frequency and amplitude on $\omega\in[0,60]$ ($B=11.5$) or $[0,30]$ ($B=14$, $17$), output bin $0.005$, and the symmetrized kernel $K(\tau,\omega)=[e^{-\tau\omega}+e^{-(\beta-\tau)\omega}]/(1+e^{-\beta\omega})$. This kernel is exact because both correlators are Hermitian autocorrelations and even under $\tau\to\beta-\tau$, so that detailed balance holds exactly and the positive-frequency transverse spectrum contains the spin-raising and spin-lowering excitations on an equal footing. The $\delta$ functions are left free because nothing in these spectra is known in advance to be sharp, and parametrizations with a built-in leading edge or quasiparticle pole~\cite{Yang2025,Yang2026} would impose such a feature where it may not exist. Grid points with relative error below $20\%$ are kept, leaving $N_\tau=15$--$61$ points per momentum (Appendix~\ref{app:sac}); the shortest windows include the $\Szz$ columns at $B=14$, so the two-magnon edges of Sec.~\ref{sec:IS2} are among the least well constrained quantities we quote. The covariance matrix comes from $40\,000$ bootstrap samples, and the sampling temperature is fixed by the criterion $a=(\langle\chi^2\rangle-\chi^2_{\min})/\sqrt{2\chi^2_{\min}}\simeq0.5$~\cite{Shao2023}, which cross-validation against held-out imaginary-time data supports for spectra of this kind~\cite{Schumm2024}; every spectrum reached $\chi^2_{\min}/N_\tau=0.21$--$1.60$ with $a=0.40$--$0.55$.

The plotted spectrum $S(\qv,\omega)$ is normalized as
\begin{equation}
  G(\qv,\tau)=\frac{1}{\pi}\int_0^\infty\! d\omega\,S(\qv,\omega)
              \bigl[e^{-\tau\omega}+e^{-(\beta-\tau)\omega}\bigr],
  \label{eq:norm}
\end{equation}
which at $\tau=0$ is the sum rule
\begin{equation}
  \int_0^\infty S(\qv,\omega)\bigl(1+e^{-\beta\omega}\bigr)\,d\omega=\pi\,G(\qv,0).
  \label{eq:sumrule}
\end{equation}
The back-transform of every one of the 159 continued spectra of Fig.~\ref{fig:maps} reproduces the measured $G(\qv,\tau)/G(\qv,0)$ to better than $5\times10^{-4}$ at every grid point and $G(\qv,0)$ to better than $3\times10^{-4}$ (Appendix~\ref{app:sac}). Where a spectrum has no weight at $\omega\lesssim1/\beta$, Eq.~\eqref{eq:sumrule} reduces to $\int_0^\infty S\,d\omega=\pi S(\qv)$, with $S(\qv)=G(\qv,0)$ the equal-time structure factor; the ratio $\int S\,d\omega/[\pi S(\qv)]$ is $1.0000$ at every momentum except the three rotor-tower columns, where thermal de-excitation weight lies at $\omega<0$ (Sec.~\ref{sec:checks}). The ``weight'' of a spectral feature means its contribution to $S(\qv)$, i.e., its share of the plotted spectrum divided by $\pi$ and corrected for thermal occupation where that matters.

Peak positions are local maxima of the spectrum smoothed with a Gaussian of width $\sigma_\omega=0.05$, which serves only to locate maxima and minima; the lower edge of a spectrum is the energy below which $0.5\%$ of its weight lies, and weight fractions are integrals of the unsmoothed spectrum between the minima bounding a peak (Appendix~\ref{app:sac}). The velocity of a gapless branch is its slope at $\delta\qv\to0$, obtained from the lowest peak $\omega_1$ at the first two momenta on the approach to the point that the branch reaches ($|\delta\qv|=\pi/8$ and $\pi/4$ along an axis, $\sqrt2$ times that along the diagonal) with the form $\omega_1=v|\delta\qv|(1-a|\delta\qv|^2)$; for $\Sxx$ the rotor offset $\Drot$ is first subtracted from $\omega_1$ (Sec.~\ref{sec:checks}). The first-step secant $\omega_1/|\delta\qv|$, a lower bound on $v$, is quoted alongside. No statistical uncertainty is attached to a peak position, the spectra being single continuations of bootstrapped correlators; where a check is available it is quoted instead.

\section{Results}
\label{sec:results}

\subsection{Overview}
\label{sec:overview}

\begin{figure*}[t]
\centering
\includegraphics[width=\textwidth]{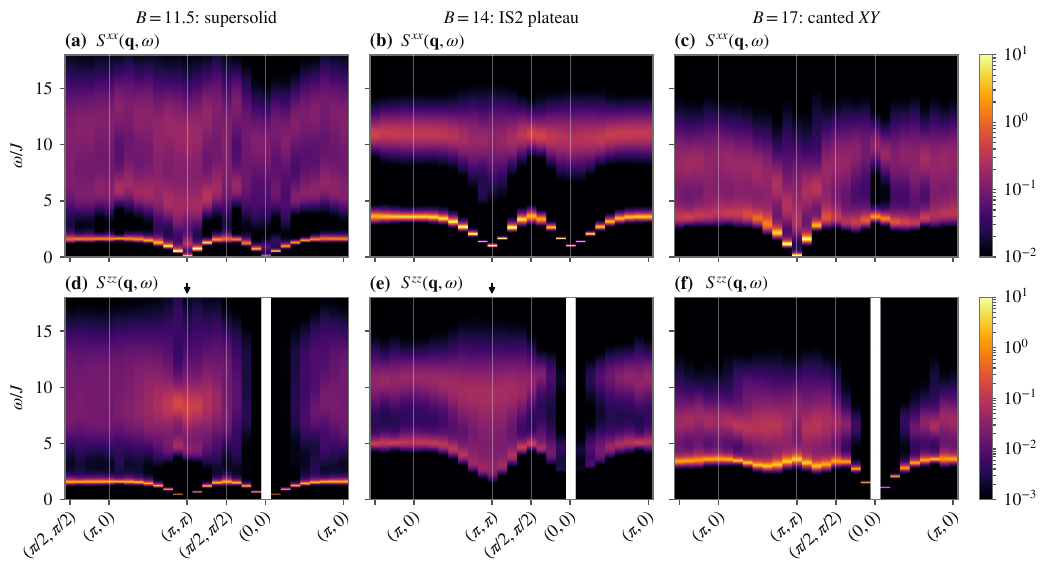}
\caption{Dynamic spin structure factors of the $S=1$ model, Eq.~\eqref{eq:H}, at $D=3.0$, $J_z=3.6$, $L=16$, $\beta=32$, along the path $(\tfrac\pi2,\tfrac\pi2)\to(\pi,0)\to(\pi,\pi)\to(\tfrac\pi2,\tfrac\pi2)\to(0,0)\to(\pi,0)$ (29 momenta). (a)--(c) Transverse $\Sxx(\qv,\omega)$ and (d)--(f) longitudinal $\Szz(\qv,\omega)$ in the supersolid ($B=11.5$), on the IS2 plateau ($B=14$), and in the canted $XY$ phase ($B=17$). Logarithmic intensity on the native $\Delta\omega=0.005$ grid of the continuation; the scale is common to the panels of one row and runs from $10^{-2}$ (transverse) and $10^{-3}$ (longitudinal) to $10^{1}$. The white column at $(0,0)$ in $\Szz$ is the purely elastic response (weight $Nm_z^2$) of the conserved $S^z_{\rm tot}$, which has no inelastic part to continue. Arrows in (d) and (e) mark the connected $(\pi,\pi)$ columns, from which the Ising-doublet weight [$>99.7\%$ of $\Szz(\Qv)$] has been removed before continuation (Sec.~\ref{sec:sectors}); in $\Sxx$ the raw correlator is continued and the saturated columns at $(\pi,\pi)$ [(a) and (c)] and $(0,0)$ [(a)] are the rotor tower of the $U(1)$ condensate.}
\label{fig:maps}
\end{figure*}

\begin{figure*}[t]
\centering
\includegraphics[width=\textwidth]{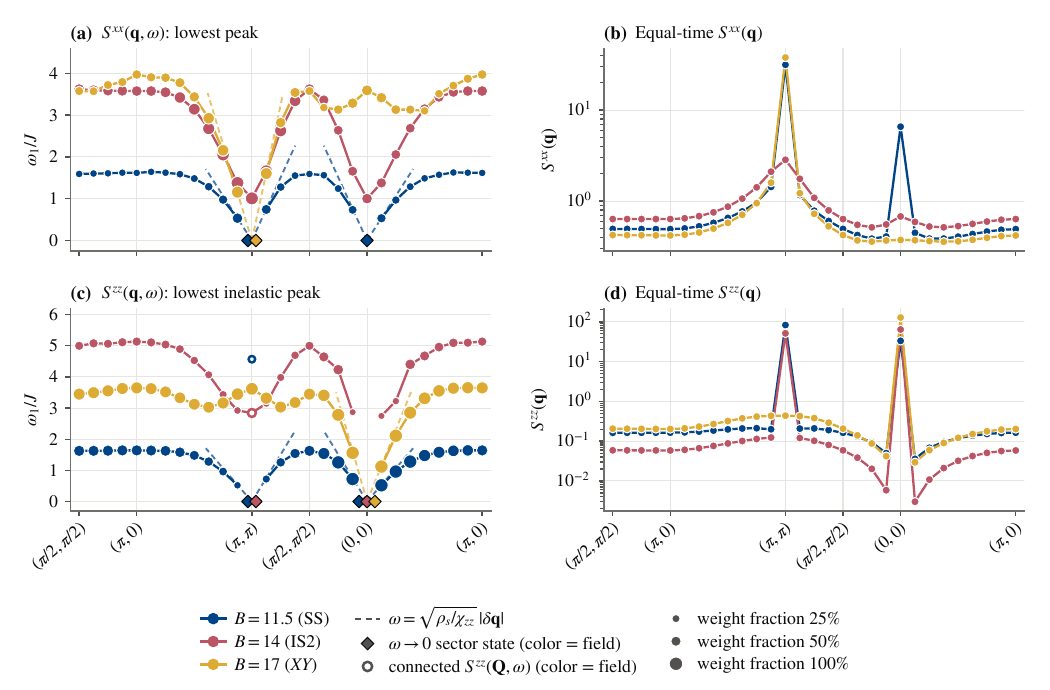}
\caption{Dispersions and static weights along the path of Fig.~\ref{fig:maps}. (a), (c) Position of the lowest peak $\omega_1$ of $S(\qv,\omega)$ (marker area growing linearly with the fraction of the total weight in that peak, key at the bottom); lines join neighboring momenta only where both carry a peak. Diamonds at $\omega=0$ (offset horizontally) mark the finite-size symmetry-sector states: in (a) the rotor tower at $(\pi,\pi)$ [SS, $XY$] and $(0,0)$ [SS]; in (c) the exact elastic line of weight $Nm_z^2$ at $(0,0)$ (all fields) and the Ising doublet at $(\pi,\pi)$ [SS, IS2]. Dashed lines: the hydrodynamic prediction $\omega=\sqrt{\rhos/\chizz}\,|\delta\qv|$ from the measured statics (Table~\ref{tab:static}), not a fit. The open circles in (c) are the lowest inelastic peaks of the connected $\Szz(\Qv,\omega)$ in the supersolid and on the plateau; the supersolid one is a weak feature carrying $8\%$ of the connected weight, so the text characterizes that spectrum by its edge $3.9$ and centroid $8.5$ instead. (b), (d) Equal-time structure factors (log scale); the $(0,0)$ values and the $(\pi,\pi)$ [SS, IS2] values in (d), and the $(\pi,\pi)$ [SS, $XY$] and $(0,0)$ [SS] values in (b), are the order-parameter weights. Statistical errors in (b) and (d) are smaller than the symbols; the peak positions in (a) and (c) carry no error estimate (Sec.~\ref{sec:sac}).}
\label{fig:summary}
\end{figure*}

\begin{figure*}[t]
\centering
\includegraphics[width=\textwidth]{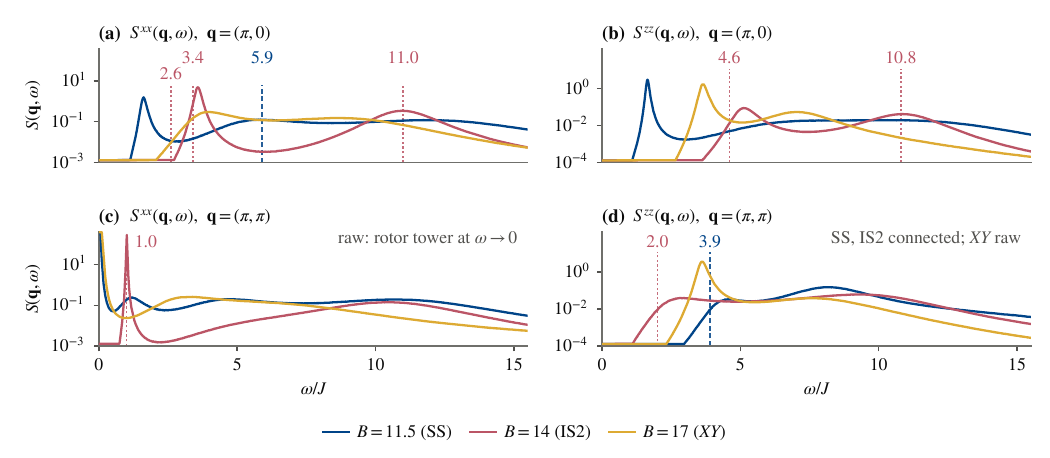}
\caption{Spectral line cuts at $(\pi,0)$ (top) and $(\pi,\pi)$ (bottom), transverse channel (left) and longitudinal channel (right), the three fields overlaid (logarithmic vertical scale, raw SAC output on the $\Delta\omega=0.005$ grid without smoothing, same normalization as Fig.~\ref{fig:maps}; the transverse column spans $10^{-3}$ to $4\times10^{2}$ and the longitudinal column $10^{-4}$ to $1.5\times10^{2}$). Dashed vertical lines mark the reference energies discussed in the text, labeled by value, one dash pattern and color per field: in (a) the local flip costs $D-B+4J_z=5.9$ (SS) and $D+B-4J_z=2.6$, $D-B+4J_z=3.4$, $B-D=11.0$ (IS2); in (b) the IS2 two-magnon threshold $\varepsilon(0)+\varepsilon(\pi,0)=4.6$ ($\varepsilon$: the lower IS2 $\Sxx$ band, Fig.~\ref{fig:is2}) and the nearest-neighbor swap cost $3J_z=10.8$; in (c) the IS2 gap $\DIS=1.0$; in (d) $2\DIS=2.0$ and the supersolid connected edge $3.9$. In (c) the $\Sxx$ spectra are the raw ones (SS and $XY$: the rotor tower saturates the scale at $\omega\to0$; IS2: gapped magnon), while in (d) the $\Szz$ spectra are the connected ones for SS and IS2 and the raw one for $XY$.}
\label{fig:cuts}
\end{figure*}

Figure~\ref{fig:maps} shows the six spectra along a path through the zone, Fig.~\ref{fig:summary} the lowest-peak dispersions and the equal-time weights, and Fig.~\ref{fig:cuts} the spectral shapes at $(\pi,0)$ and $\Qv$. The equal-time weights that normalize each column already identify the phases (Table~\ref{tab:static}). The plateau has $m_z=0.500$ to three digits and $\Szz(\Qv)/N=0.20$, the checkerboard $|{+1},0\rangle$ solid ($0.25$ classically) with quantum reduction. The supersolid carries a folded condensate peak $\Sxx(0,0)/N=0.026$, $21\%$ of its $\Qv$ peak and an order of magnitude above the other two phases; this absolute weight, not the ratio to the $\Qv$ peak (similar in IS2 only because both weights are small there), is the two-peak diagnostic of Fig.~1(e) of the Letter. The plateau is incompressible ($\chizz\simeq0$, $\rhos\simeq0$), while the supersolid and the $XY$ phase are superfluids, the supersolid being $2.7$ times more compressible and $1.7$ times less stiff. Every number agrees with Fig.~1 of the Letter once its fields and its $\rhos$ are converted to the units of Eq.~\eqref{eq:H} (Sec.~\ref{sec:params}). The last two rows of Table~\ref{tab:static} are the hydrodynamic predictions tested in Sec.~\ref{sec:checks}.

In two dimensions at $T>0$ the $U(1)$ order is quasi-long-range in the Berezinskii-Kosterlitz-Thouless sense~\cite{Kosterlitz1973}, and ``condensate'' and ``Goldstone mode'' are meant in that sense. The measured stiffness exceeds the universal-jump value $2T/\pi=0.020$~\cite{Nelson1977} by factors of $13$ (supersolid) and $22$ ($XY$), so the transverse correlations decay with exponent $\eta=T/(2\pi\rhos)\lesssim0.02$, by a few percent across $L=16$, and the finite system behaves as ordered. The $Z_2$ checkerboard orders at finite temperature in the usual sense.

\subsection{Finite-size states at $\Qv$ and $(0,0)$}
\label{sec:sectors}

\begin{table*}[t]
\caption{Where the gapless (Goldstone) branch is seen. ``yes'' means a peak dispersing linearly to zero, with the velocity defined in Sec.~\ref{sec:sac} ($\delta\qv\to0$ slope from the first two axial momenta, after subtracting the rotor offset $\Drot$ from the $\Sxx$ peaks; resolution $0.01$); the first-step secants are $1.32$ ($\Sxx$) and $1.33$ ($\Szz$) in the SS and $2.85$ and $2.86$ in the $XY$ phase. ``Gapped'' gives the lowest-peak energy at that point. Exactly at $\Qv$, $\Szz$ is the Ising doublet plus a gapped connected remainder in the SS and IS2 (edges $3.9$ and $2.0$) and an ordinary gapped mode in the $XY$ phase.}
\label{tab:goldstone}
\begin{ruledtabular}
\begin{tabular}{lcccc}
 & $\Sxx$ near $\Qv$ & $\Sxx$ near $(0,0)$ & $\Szz$ near $(0,0)$ & $\Szz$ near $\Qv$\\
\hline
SS ($B=11.5$)  & yes, $v=1.35$ & yes, $v=1.35$ & yes, $v=1.37$ & yes, $v=1.37$\\
IS2 ($B=14$)   & gapped 1.00   & gapped 1.00   & two-magnon, $2\DIS=2.0$\footnotemark[1] & two-magnon, $2\DIS=2.0$\footnotemark[1]\\
$XY$ ($B=17$)  & yes, $v=2.90$ & gapped 3.59   & yes, $v=2.91$ & gapped 3.62\\
\end{tabular}
\end{ruledtabular}
\footnotetext[1]{Computed threshold at the point itself (Fig.~\ref{fig:is2}), a lower bound on the onset (Sec.~\ref{sec:IS2}), not a measured peak; the nearest measured $\Szz$ edges are $2.18$ at $(\tfrac\pi8,0)$ and $1.98$ at $\Qv$.}
\end{table*}

On a finite periodic lattice $\mathcal H$ is translation invariant and $S^z_{\rm tot}$ is conserved, so no eigenstate breaks the $Z_2$ sublattice symmetry or the $U(1)$ spin rotation. The ordered ground state is a small set of quasi-degenerate symmetric eigenstates, and $S^z_\qv$ and $S^x_\qv$ at the ordering wave vectors connect the ground state to its partners in that set~\cite{Anderson1952,Bernu1992,Lhuillier2005,Tasaki2019}. This decides how the columns at $\Qv$ and $(0,0)$ are treated. The simplest is $\Szz$ at $\qv=(0,0)$: $S^z_{\qv=0}$ commutes with $\mathcal H$, so $\Gzz(\qv=0,\tau)=Nm_z^2$ in every phase and the column is a purely elastic line, with no inelastic spectrum to continue.

\paragraph{Ising doublet in $\Szz(\Qv,\omega)$.} The checkerboard order parameter is $M_{\rm stag}=\sum_i(-1)^{x_i+y_i}S^z_i=\sqrt N S^z_\Qv$, with $m_{\rm stag}=M_{\rm stag}/N$. Translation by one lattice spacing maps the pattern $|A\rangle$ onto $|B\rangle$, and the eigenstates $|\pm\rangle=(|A\rangle\pm|B\rangle)/\sqrt2$, of crystal momentum $0$ and $\Qv$, are split by a tunneling amplitude $\Delta_{\rm dbl}\sim e^{-cL^2}$: going from $|A\rangle$ to $|B\rangle$ takes a domain wall in imaginary time that covers the whole $L\times L$ lattice. $S^z_\Qv$ connects them with the macroscopic matrix element $\sqrt N m_{\rm stag}$, so that
\begin{equation}
  \Gzz(\Qv,\tau)\simeq Nm_{\rm stag}^2\,\frac{\cosh[\Delta_{\rm dbl}(\beta/2-\tau)]}{\cosh(\beta\Delta_{\rm dbl}/2)}+\text{inelastic},
\end{equation}
a line at $\omega=\pm\Delta_{\rm dbl}$ of weight $Nm_{\rm stag}^2$ whose $\tau$ dependence for $\beta\Delta_{\rm dbl}\ll1$, $1-\tfrac12\Delta_{\rm dbl}^2\tau(\beta-\tau)$, is quadratic in $\Delta_{\rm dbl}$. In the supersolid $\Gzz(\Qv,\tau)$ is flat from $\tau=2$ to the end of the grid at $\tau=15.6$ to one part in $10^7$, against a statistical error of $2\times10^{-8}$ on $\Gzz(\Qv,\tau)/\Gzz(\Qv,0)$, which bounds $\Delta_{\rm dbl}\lesssim3\times10^{-5}$. The doublet holds more than $99.7\%$ of $\Szz(\Qv)=82.8$ (SS) and $51.2$ (IS2), i.e., $m_{\rm stag}^2=0.32$ and $0.20$. A raw continuation cannot represent this exactly elastic line (the fit stalls at $\chi^2/N_\tau\simeq200$ in the SS and $\simeq10^8$ in IS2), so for these two columns [arrows in Figs.~\ref{fig:maps}(d) and \ref{fig:maps}(e)] we continue the connected correlator, obtained by subtracting the plateau at $\tau\simeq\beta/2$ per bootstrap sample (Bragg subtraction). What remains of $\Szz(\Qv)$ after the subtraction, the connected weight $\Szz(\Qv)-\Gzz(\Qv,\beta/2)$, is $0.164$ (SS) and $0.127$ (IS2), and any inelastic weight above $\omega\simeq1/\beta$ removed with the plateau is below $0.1\%$ of it (Appendix~\ref{app:sac}). In the $XY$ phase there is no solid order and the raw correlator is used.

\paragraph{Rotor tower in $\Sxx$ at $\Qv$ and $(0,0)$.} For the broken $U(1)$ symmetry the symmetric eigenstates form a tower rather than a doublet~\cite{Anderson1952,Bernu1992}. The condensate with in-plane moment $m_\perp$ is a wave packet of total-$S^z$ sectors around $Nm_z$, with rotor energies
\begin{equation}
  E(S^z_{\rm tot})-E_{\rm min}\simeq\frac{(S^z_{\rm tot}-Nm_z)^2}{2\chizz N},\qquad
  \Drot\simeq\frac{1}{2\chizz N},
  \label{eq:rotor}
\end{equation}
since the moment of inertia of the rotor is $\chizz N$; the gap is a power of $1/N$ rather than exponentially small. $S^x_\Qv$ changes $S^z_{\rm tot}$ by $\pm1$ and connects the ground state to the two adjacent rotor levels with total weight $\simeq\tfrac12Nm_\perp^2$,
\begin{equation}
  \Gxx(\Qv,\tau)\simeq\tfrac12Nm_\perp^2\,e^{-\Drot\tau}+\text{inelastic}
\end{equation}
at $T=0$, i.e., a line at $\omega=\Drot$ carrying the condensate weight. With the measured $\chizz$ (Table~\ref{tab:static}), $\Drot=0.014$ in the supersolid and $0.038$ in the $XY$ phase, below the resolution at $\beta=32$; the continuation puts essentially all of this weight into the lowest bins ($\omega\simeq0.03$), the saturated columns of Figs.~\ref{fig:maps}(a) and \ref{fig:maps}(c) and the diamonds of Fig.~\ref{fig:summary}(a). In the supersolid the checkerboard folds $\Qv$ onto $(0,0)$, so the same tower appears in $\Sxx$ at $(0,0)$. Because $\beta\Drot=0.46$ and $1.2$, the tower is thermally populated and $\Gxx(\Qv,\tau)$ keeps a $\tau$ dependence, so subtracting its $\beta/2$ value is not meaningful (the attempt does not converge, $\chi^2/N_\tau\simeq2\times10^5$ at $B=17$) and the raw correlator is continued; Sec.~\ref{sec:checks} tests the tower quantitatively. Both sectors become elastic Bragg peaks as $L\to\infty$~\cite{Penrose1956}. The Goldstone mode is the separate branch that disperses \emph{toward} these points, and it is what we mean by ``gapless'' below.

\subsection{One Goldstone mode in four places}
\label{sec:doubling}

Table~\ref{tab:goldstone}, which collects the results of Secs.~\ref{sec:SS}--\ref{sec:XY}, is the central result. The supersolid has one Goldstone mode, and because the $Z_2$ checkerboard halves the Brillouin zone that mode is visible on the approach to both $\Qv$ and $(0,0)$, in both channels, with a single velocity $v=1.36\pm0.01$. Each neighbor shows only part of the pattern: the $XY$ superfluid has the mode at $\Qv$ (transverse) and at $(0,0)$ (longitudinal) only, and the IS2 solid has none. The counting is the one of Sec.~\ref{sec:intro}: one broken generator, hence one type-A mode, linear despite the finite magnetization. Generalized spin-wave theory of the related dimer model finds the same doubling in the longitudinal channel~\cite{Murakami2013}.

The doubling is the dynamical counterpart of the two-peak $\SpmPRL(\qv)$ that the Letter used as its supersolid diagnostic, and as detections of the folding the two are close to equivalent, since a uniform in-plane component on a square lattice itself requires sublattice inequivalence. What the dynamics adds is quantitative: the branches at $\Qv$ and $(0,0)$ share one velocity, and that velocity and the sound-mode weight match the static $\rhos$ and $\chizz$ (Sec.~\ref{sec:checks}). This distinguishes a homogeneous supersolid from a sample separated into solid and superfluid regions, which is what hard-core bosons do when doped away from the checkerboard~\cite{Batrouni2000,Sengupta2005}: there the mode velocity would be that of the superfluid regions while $\rhos$ and $\chizz$ are sample averages, so the two need not agree, whereas here they agree to within $1\%$ in both superfluid phases. Phase separation is in any case not expected at $B=11.5$, which lies between two continuous transitions; the Letter places it at the first-order IS2--$XY$ boundary.

\subsection{Supersolid, $B=11.5$}
\label{sec:SS}

\paragraph{Transverse channel.} A sharp, intense low-energy branch runs through the whole zone [Figs.~\ref{fig:maps}(a) and \ref{fig:summary}(a)]. Along the zone boundary $|q_x|+|q_y|=\pi$ of the checkerboard (the first segment of the path) and on to $(\pi,3\pi/8)$ it is flat at $\omega_1\simeq1.6$ and carries $21$--$24\%$ of the local weight. It disperses linearly to zero on the approach to $\Qv$ and to $(0,0)$, its weight fraction rising to $58\%$ next to $\Qv$ as the peak sharpens, and both ``V''s have the slope $v=1.35$--$1.37$ at $\delta\qv\to0$ (Sec.~\ref{sec:checks}; the first-step secants are $1.32$ along the axes and $1.29$--$1.31$ along the diagonal). The columns at $\Qv$ and $(0,0)$ themselves carry the rotor tower (Sec.~\ref{sec:sectors}), with $\Sxx(0,0)=6.6$, fifteen times the neighboring values. Above the branch, a broad continuum from $\omega\simeq4.5$ to $14$ holds $\sim78\%$ of the weight far from $\Qv$, with components at $5.7$--$6.5$ and $10$--$13$. The first matches the cost $D-B+4J_z=5.9$ of raising an $S^z=0$ site of the depleted sublattice, surrounded by $S^z=+1$ neighbors, to $S^z=+1$ [marked in Fig.~\ref{fig:cuts}(a)], while the flips $-1\to0$ and $0\to-1$ cost only $\mp0.1$: these soft flips form the condensate, and their hopping band (Sec.~\ref{sec:phases}), not the $0.1$ cost, sets the scale of the Goldstone branch, whose flat top is measured at $1.6$. The dark window $2\lesssim\omega\lesssim4.5$ has little weight but is not a gap: two-magnon states, a magnon paired with a soft one at $\Qv$ or $(0,0)$, extend down to the branch itself but carry little transverse weight.

\paragraph{Longitudinal channel.} Over most of the zone $\Szz$ is a single sharp mode [Figs.~\ref{fig:maps}(d) and \ref{fig:summary}(c)]. Along the zone boundary it sits at $\omega_1=1.63$--$1.64$ with $67$--$68\%$ of the weight (the rest in a weak continuum at $\omega\simeq8$--$10$), at the energy of the flat top of the transverse branch: in a state with both $m_z\ne0$ and a canted condensate the transverse magnon has a longitudinal component (Sec.~\ref{sec:XY}), so the same quasiparticle is visible in $\Szz$. Approaching $(0,0)$ the mode softens linearly with $v=1.36$--$1.37$ (first-step secant $1.33$) and is the only resolved feature ($100\%$ of the weight): this is the sound (phase) mode of the spin superfluid, seen through the density conjugate to the phase, and the large-$\tau$ decay of $\Gzz$ confirms it without any continuation (Sec.~\ref{sec:checks}). Approaching $\Qv$ the same soft branch appears again, now with $21$--$45\%$ of the weight, least ($21\%$) next to $\Qv$, the rest moving into the bright lobe at $\omega\simeq8$--$9$ centered on $\Qv$ in Fig.~\ref{fig:maps}(d). At $\Qv$ exactly the response is the Ising doublet plus a connected remainder with no resolvable low-energy weight: its lower edge is $3.9$ ($4.2$ in an independent 1152-bin data set; Appendix~\ref{app:sac}) and centroid $8.5$ [Fig.~\ref{fig:cuts}(d)], with no resolved substructure, since the connected signal exceeds its noise only out to $\tau\simeq2$. The dark dome below $\omega\simeq4$ around $\Qv$ in Fig.~\ref{fig:maps}(d) is this gap: fluctuations of the checkerboard amplitude cost energy, and the $Z_2$ solid has no Goldstone mode.

That $\Szz$ is gapless at all, although the supersolid breaks the same $Z_2$ symmetry as IS2, is a consequence of compressibility. The gap of IS2 comes not from the $Z_2$ order but from the absence of a superfluid component: IS2 has $\chizz=0$, so changing the density anywhere costs the particle-hole gap. The $Z_2$ order by itself gaps only the amplitude fluctuations at $\Qv$, and the supersolid inherits that (edge $3.9$ against $2.0$ in IS2). But the supersolid is compressible and the density $S^z_\qv$ at small $\qv$ is conjugate to the condensate phase, so hydrodynamics requires a sound mode $\omega=v|\qv|$ with weight $\propto\chizz v|\qv|$, which the checkerboard folds onto $\qv\to\Qv$; a spectrum gapped at small $\qv$ with vanishing weight could not keep the compressibility $\chizz=\lim_{\qv\to0}(2/\pi)\int_0^\infty\Szz(\qv,\omega)\,\omega^{-1}d\omega$ finite. At $\Qv$ exactly the supersolid thus looks like IS2, on the approach to $(0,0)$ like the $XY$ phase, and on the approach to $\Qv$ it shows the folded copy of the sound mode, which neither neighbor has.

\paragraph{Sublattice structure.} The longitudinal weight is small [$\Szz(\qv)\simeq0.04$--$0.21$ against $\Sxx\simeq0.4$--$1.4$] because the $\Qv$ Bragg peak takes $53\%$ of the zone sum $\langle(S^z_i)^2\rangle=0.607$. The two measured order parameters fix the moments of each sublattice. With $m_z=0.3597$ and $m_{\rm stag}=\sqrt{\Szz(\Qv)/N}=0.5686$, $\langle S^z\rangle=m_z\pm m_{\rm stag}$ is $+0.928$ on sublattice $A$ and $-0.209$ on $B$, with $A$ and $B$ labeled as in the plateau state $|{+1}\rangle_A|0\rangle_B$ of Sec.~\ref{sec:IS2}. Since $\langle S^z\rangle=p_{+1}-p_{-1}$ and $\langle(S^z)^2\rangle=p_{+1}+p_{-1}$ for occupations $p_m$, $\langle(S^z)^2\rangle$ lies between $0.928$ and $1$ on $A$, and the zone sum, the average of the two sublattices ($2\times0.607=1.214$ in total), leaves $0.21$--$0.29$ on $B$: $71$--$79\%$ of the $B$ sites are in $|0\rangle$ and $21$--$25\%$ in $|{-1}\rangle$. For the staggered in-plane pattern $\langle S^+\rangle=m_Ae^{i\phi}$ on $A$ and $-m_Be^{i\phi}$ on $B$, $\Sxx(\Qv)/N=\tfrac12[(m_A+m_B)/2]^2$ and $\Sxx(0,0)/N=\tfrac12[(m_A-m_B)/2]^2$, so the folded condensate at $(0,0)$ measures the difference of the two moments. The measured $0.123$ and $0.026$ give in-plane moments $0.72$ and $0.27$, and since the nearly saturated $A$ sublattice can hold at most $\sqrt{1-0.928^2}=0.37$ in the plane, $m_B=0.72$ and $m_A=0.27$. The condensate is therefore not a dilute gas of defects: $A$ is essentially saturated and inert, and the strongly doped $B$ sublattice carries the condensate, which is the picture behind the SS--IS2 transition of Sec.~\ref{sec:IS2}.

\subsection{IS2 plateau, $B=14$}
\label{sec:IS2}

\begin{figure}[t]
\centering
\includegraphics[width=\columnwidth]{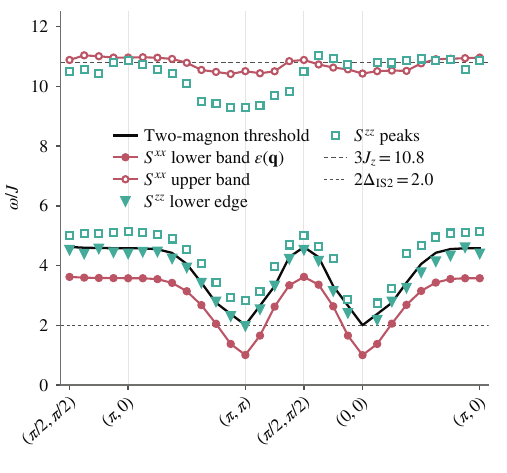}
\caption{Two-magnon threshold on the IS2 plateau, $B=14$. Filled rose circles, the lower $\Sxx$ magnon band (lowest peak); open rose circles, its upper band (tallest peak above $\omega=8$); black line, the two-magnon threshold $\min_\kv[\varepsilon(\kv)+\varepsilon(\qv-\kv)]$ computed from the lower band on the full zone (81 momenta extended by reflection); teal triangles, the lower edge of $\Szz(\qv,\omega)$; open teal squares, its peaks. At $(\pi,\pi)$ the $\Szz$ points are from the connected spectrum. The measured edge lies below the computed threshold by $0.12$ on average (Sec.~\ref{sec:IS2}). Horizontal dashed lines: $2\DIS=2.0$ and the nearest-neighbor swap cost $3J_z=10.8$.}
\label{fig:is2}
\end{figure}

\paragraph{Transverse channel.} The spectrum is that of a gapped Ising solid [Figs.~\ref{fig:maps}(b) and \ref{fig:cuts}(c)], with no gapless mode anywhere, and there cannot be one: the plateau breaks only the discrete $Z_2$ symmetry, the $U(1)$ symmetry being intact [$\Sxx(\Qv)/N=0.011$, extrapolating to zero with $N$ in the Letter; $\rhos\simeq0$]. A coherent magnon band has its minimum $\DIS=1.00$ at both $\Qv$ and $(0,0)$, which are equivalent because the checkerboard doubles the unit cell, and rises to $3.58$--$3.62$ at $(\pi,0)$ and $(\tfrac\pi2,\tfrac\pi2)$, carrying $87$--$92\%$ of the weight at and next to $\Qv$ and $48$--$84\%$ elsewhere. It stems from the two flips of a $B$ site of the plateau state $|{+1}\rangle_A|0\rangle_B$, raising it at cost $D-B+4J_z=3.4$ and lowering it at $D+B-4J_z=2.6$ [Fig.~\ref{fig:cuts}(a)], which hopping spreads into a band; since $\Sxx$ superposes the $S^+$ and $S^-$ flavors, whose bare levels are only $0.8$ apart while the band is $2.6$ wide, what is seen is their lower envelope. Lowering an $A$ site costs $B-D=11.0$ and, its neighbors being $S^z=0$, barely disperses: this is the second, nearly flat band at $\omega=10.4$--$11.0$, with $8$--$52\%$ of the column.

The gap measures the distance to the lower plateau edge. The plateau ground state is field independent, being an eigenstate at fixed $S^z_{\rm tot}=N/2$, so an excitation with $\Delta S^z=-1$ has $\DIS(B)=B-B_c$ exactly, where $B_c$ is the field at which it condenses. The measured $\DIS=1.00$ at $B=14$ therefore gives $B_c=13.0$ from a single field, which agrees, within the $0.05$ to which a peak position is stable (Appendix~\ref{app:sac}), with $B_c=13.01\pm0.01$ from an $L=16$ field scan (Appendix~\ref{app:scan}). This assumes that the cheaper flavor, which the local costs identify as the hole ($2.6$ against $3.4$), stays lowest across the plateau. The supersolid Goldstone mode is thus born from this hole band, which softens to zero at $\Qv$ at the continuous SS--IS2 transition. On the high-field side the plateau ends in the first-order transition at $B^*=14.81$ (Appendix~\ref{app:scan}), so the $XY$ Goldstone mode is not continuously connected to the plateau band.

\paragraph{Longitudinal channel.} There is no sharp mode, and away from $\Qv$ the intensity on the common scale of Fig.~\ref{fig:maps}(e) is a factor $10$ below that of the supersolid [the equal-time weight is $\Szz(\qv)=0.003$--$0.12$]. The spectrum is a broad continuum whose lower edge is $2.0$ at $\Qv$ [connected spectrum, Fig.~\ref{fig:cuts}(d)] and rises to $4.4$--$4.6$ along the zone boundary [Fig.~\ref{fig:cuts}(b)]. A longitudinal probe of a gapped Ising state creates pairs of transverse magnons, and Fig.~\ref{fig:is2} makes this quantitative: the two-magnon threshold $\min_\kv[\varepsilon(\kv)+\varepsilon(\qv-\kv)]$, computed from the lower $\Sxx$ band extracted at all 81 momenta and a lower bound on the particle-hole onset (see below), has the minima $2\DIS=2.0$ at $\Qv$ and $(0,0)$ and stays near $4.6$ along the zone boundary [$\varepsilon(0)+\varepsilon(\pi,0)=4.6$ at $(\pi,0)$], and the measured edge tracks its momentum dependence over the whole path. The upper lobe at $\omega\simeq9$--$11$ matches the simplest $S^z_{\rm tot}$-conserving process, a nearest-neighbor swap $|{+1}\rangle_A|0\rangle_B\to|0\rangle_A|{+1}\rangle_B$ that gives the moved $+1$ three $+1$ neighbors and costs $3J_z=10.8$. It lies below the free two-band continuum ($\ge11.4$), so at high energy the probe excites a bound pair rather than two free magnons.

The measured edge lies slightly \emph{below} the threshold, by $0.12$ on average over the 26 distinct momenta of the path other than $(0,0)$ (maximum $0.29$; above it at only four points, by at most $0.04$). A finite frequency resolution smears a sharp onset in both directions, and the $0.5\%$-quantile edge picks up the part smeared downward; if the true onset is the computed threshold, this offset therefore measures the resolution with which an edge can be located here. Since $S^z_\qv$ conserves $S^z_{\rm tot}$, however, the pair is one particle and one hole, so the threshold computed from the lower envelope is only a lower bound on the true onset, reached if both flavors share one minimum. The bare costs $2.6$ and $3.4$ suggest they may not, in which case the true onset at $\Qv$ lies nearer the $2.8$ peak of the connected spectrum and the measured edge is the low-weight tail of the continuum. Either way, the edge is not evidence for states below the continuum.

\subsection{Canted $XY$ antiferromagnet, $B=17$}
\label{sec:XY}

\paragraph{Transverse channel.} A linear Goldstone mode emanates from $\Qv$ only [Fig.~\ref{fig:maps}(c)], with $v=2.90$ (Sec.~\ref{sec:checks}; first-step secants $2.85$ along the axis and $2.82$ along the diagonal), more than twice the supersolid value; the $\Qv$ column is the rotor tower of the $XY$ condensate. $(0,0)$ is fully gapped, $\omega_1=3.59$ with $65\%$ of the weight, and $\Sxx(0,0)=0.375$ shows no enhancement over its neighbors: with no solid order there is nothing to fold $\Qv$ onto $(0,0)$, and the two-peak structure of the supersolid is gone, as in the Letter. Away from $\Qv$ and $(0,0)$ the lowest band, at $\omega_1\simeq3.1$--$4.0$, is much less coherent than on the plateau, with only $32$--$49\%$ of the weight and the rest in a broad continuum at $6$--$10$, consistent with the coupling of one- and two-magnon states that noncollinear canted order allows~\cite{Zhitomirsky2013}.

\paragraph{Longitudinal channel.} A sharp single mode ($73$--$90\%$ of the weight along $(\pi,0)$--$(\pi,\pi)$, $100\%$ at small $q$) sits at $\omega_1\simeq3.0$--$3.65$ [Fig.~\ref{fig:maps}(f)] and softens linearly toward $(0,0)$ with $v=2.91$ (first-step secants $2.86$ and $2.81$), the velocity of the transverse Goldstone mode at $\Qv$. This is the standard result for a $U(1)$ spin superfluid: $S^z_\qv$ is conjugate to the condensate phase, so the phase mode appears in $\Szz$ at $\qv\to0$ as a sound mode of small weight, $\Szz(\qv)\propto q$. In linear spin-wave theory of the canted state~\cite{Zhitomirsky1998} the lab-frame $S^z_i$ of a spin tilted by $\theta$ from $z$ contains $-\sin\theta\,(-1)^{x_i+y_i}$ times its local transverse component, so $S^z_\qv$ at small $\qv$ creates the Goldstone state at $\Qv+\qv$. The same decomposition places the other, gapped one-magnon branch in $\Szz$ near $\Qv$ and in $\Sxx$ near $(0,0)$, which is why the two gapped values agree ($3.62$ and $3.59$, within the $0.05$ stability of a peak position): the two channels show the same two branches, swapped between $(0,0)$ and $\Qv$.

\subsection{Three parameter-free checks}
\label{sec:checks}

\begin{figure*}[t]
\centering
\includegraphics[width=\textwidth]{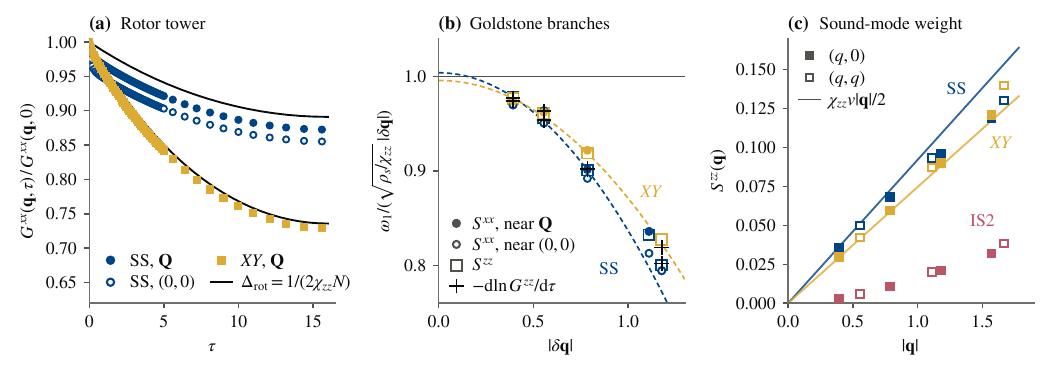}
\caption{Parameter-free hydrodynamic checks, all using the measured $\rhos$ and $\chizz$ of Table~\ref{tab:static}. (a) Raw transverse correlator at $\Qv$ [and, for the supersolid, at $(0,0)$] normalized to $\tau=0$, against the thermal rigid-rotor tower of Eq.~\eqref{eq:rotorG} (solid black lines), whose only inputs are the gap $\Drot=1/(2\chizz N)$ of Table~\ref{tab:static} and $m_z$, so that one curve serves $\Qv$ and $(0,0)$; the constant offset is the inelastic weight. (b) Lowest-peak energy of every gapless branch divided by the hydrodynamic value $\sqrt{\rhos/\chizz}\,|\delta\qv|$, versus the distance $|\delta\qv|$ from the point it approaches (small circles $\Sxx$ after subtraction of the rotor offset $\Drot$, filled on the approach to $\Qv$ and open on the approach to $(0,0)$; large open squares $\Szz$, on the approach to both points, where they coincide; color by field as in (a); axial and diagonal steps). Plus signs are the continuation-free large-$\tau$ decay rates $-d\ln\Gzz/d\tau$ at the momenta listed in Sec.~\ref{sec:checks}, divided in the same way. Dashed curves: $c-a|\delta\qv|^2$ through the first two axial $\Szz$ points of each phase, with $c=1.004$ (SS) and $0.996$ ($XY$). (c) Equal-time $\Szz(\qv)$ on the approach to $(0,0)$ along $(q,0)$ and $(q,q)$ against the sound-mode weight $\chizz v|\qv|/2$, a parameter-free test at the smallest momentum of each phase; the IS2 points show what an incompressible state gives. Statistical errors of the QMC data in (a) and (c) are smaller than the symbols.}
\label{fig:hydro}
\end{figure*}

The spectra are tied to raw Monte Carlo observables by three comparisons that use only the $\rhos$ and $\chizz$ of Table~\ref{tab:static} (Fig.~\ref{fig:hydro}), and the soft longitudinal branch is confirmed without analytic continuation.

\paragraph{Velocity.} Each of the four gapless branches of the supersolid (two channels, two special points) is sampled along an axis and along the diagonal, and the $XY$ phase has two branches. The raw first-step secants span $1.30$ to $1.36$ in the supersolid and $2.81$ to $2.95$ in the $XY$ phase, and two systematic effects account for this spread. The first is the rotor tower of Sec.~\ref{sec:sectors}: the transverse peaks lie above the longitudinal ones by $0.010$--$0.020$ in the supersolid and by $0.035$--$0.040$ in the $XY$ phase. $S^x_\qv$ changes $S^z_{\rm tot}$ by $\pm1$, so a transverse excitation costs the Goldstone energy plus one level spacing of the tower [$\varepsilon^\pm_M$ of Eq.~\eqref{eq:rotorG} below], whose mean over the two signs is $\Drot=0.0144$ (SS) and $0.0383$ ($XY$) in every sector $M$, whereas $S^z_\qv$ stays within a sector. With $\Drot$ subtracted from the $\Sxx$ peaks, the axial secants are $1.32$ ($\Sxx$) and $1.33$ ($\Szz$) in the supersolid and $2.85$ and $2.86$ in the $XY$ phase, and the diagonal ones are $1.29$--$1.31$ and $2.81$--$2.82$. The two channels then agree to $1\%$ and $0.3\%$, which is the $0.005$ output grid of the spectra, and the remaining difference is between the two directions. The second effect is curvature. A secant is a lower bound on the $\delta\qv\to0$ slope of a branch that bends over toward its zone-boundary maximum, and Fig.~\ref{fig:hydro}(b) shows the bending: the ratio of the peak energy to $\sqrt{\rhos/\chizz}\,|\delta\qv|$ falls from $0.97$--$0.98$ at the first axial step to $0.89$--$0.92$ at the second and $0.79$--$0.83$ at the third, on nearly one curve for both channels and both phases. For a gapless branch the leading correction is cubic, $\omega=v|\delta\qv|(1-a|\delta\qv|^2)$, and this form through the first two momenta of a branch gives $v=1.35$--$1.37$ for all eight branches of the supersolid, axial and diagonal alike, and $2.90$ and $2.91$ for the two axial branches of the $XY$ phase [dashed curves in Fig.~\ref{fig:hydro}(b)]. We quote $v=1.36\pm0.01$ and $2.91\pm0.01$, where the uncertainty is the spread over branches and also what the output grid allows ($0.008$ in $v$). Adding the third axial momentum and a quintic term changes $v$ by at most $0.004$, and the continuation-free decay rates quoted below give $1.36$ and $2.93$ in the same way. For a $U(1)$ superfluid the phase $\phi$ and the conserved density are conjugate, with energy density $\tfrac12\rhos(\nabla\phi)^2+\tfrac12\chizz^{-1}(\delta S^z)^2$, so hydrodynamics fixes $v^2=\rhos/\chizz$~\cite{Halperin1969}: $\sqrt{\rhos/\chizz}=1.36$ in the supersolid and $2.93$ in the $XY$ phase. The measured velocities agree with these values to within $1\%$ ($1.359$ against $1.360$ and $2.908$ against $2.927$ for the means over branches), and this holds separately for the transverse branches at $\Qv$ and the longitudinal branch at $(0,0)$, which are therefore one and the same hydrodynamic mode; the first-step secants lie $2$--$3\%$ (axial) and $4$--$5\%$ (diagonal) below. The extrapolation assumes a gapless branch. A straight line through the first two axial momenta has a slope $13$--$19\%$ below $\sqrt{\rhos/\chizz}$ and a positive intercept of $0.07$ to $0.13$, so reading the intercept as a gap would contradict the hydrodynamic velocity, although two or three momenta at one size cannot exclude a small gap by themselves. The factor $2.2$ between the two phases decomposes into $\sqrt{\rhos^{XY}/\rhos^{\rm SS}}=1.32$ from the stiffness and $\sqrt{\chizz^{\rm SS}/\chizz^{XY}}=1.63$ from the compressibility, so at $B=11.5$, close to the degeneracy point $zJ_z-D=11.4$, the supersolid is softer mainly because it is more compressible. This is not a property of the phase: toward the plateau $dm_z/dB$ falls [from $0.34$ at $B=11.0$ to $0.08$ at $12.5$ in the $L=12$ scan of Fig.~\ref{fig:scan}(a)] while $\rhos$ stays at $0.15$--$0.25$, and the stiffness and compressibility factors become equal (both $1.41$) near $B=12$.

\paragraph{Sound-mode weight and continuation-free rates.} If a single sound pole at $\omega=v|\qv|$ exhausts the weight at $T=0$, the equal-time weight is the ground-state fluctuation of the density conjugate to the phase, $\Szz(\qv)=\chizz v|\qv|/2$: each Fourier mode of the phase is an oscillator of mass $\chizz$ and frequency $v|\qv|$, and the density is its momentum. With $\chizz=0.1355$ and $v=\sqrt{\rhos/\chizz}=1.36$ this predicts $0.036$, $0.072$, $0.109$ at $\qv=(\tfrac\pi8,0)$, $(\tfrac\pi4,0)$, $(\tfrac{3\pi}8,0)$ in the supersolid against the measured $0.036$, $0.068$, $0.096$, and with $\chizz=0.0510$, $v=2.93$ it predicts $0.029$, $0.059$, $0.088$ in the $XY$ phase against the measured $0.030$, $0.059$, $0.090$ [Fig.~\ref{fig:hydro}(c)]. Only the smallest momentum is a parameter-free test, since beyond it the weight involves $\chizz(\qv)$ rather than its $\qv\to0$ limit; there the supersolid agrees to $1\%$, and the $XY$ values match at all three momenta. On the plateau $\Szz(\qv)$ is an order of magnitude smaller [$0.003$ at $(\tfrac\pi8,0)$], as an incompressible state requires. The soft longitudinal branch can also be established without analytic continuation. For a sum of decaying exponentials the local log slope $-d\ln\Gzz/d\tau$ decreases monotonically toward the lowest excitation energy that carries weight, so its large-$\tau$ value is an upper bound on that energy. Fitted over the last $40\%$ of the $\tau$ range in which $\Gzz>5\sigma$ (a fit that includes the mirror term $e^{-(\beta-\tau)\omega}$ gives the same values to three digits), it is $0.52$, $0.96$ and $1.29$ at $(\tfrac\pi8,0)$, $(\tfrac\pi4,0)$ and $(\tfrac{3\pi}8,0)$ and $0.72$ at $(\tfrac\pi8,\tfrac\pi8)$ in the supersolid, the first two giving $v=1.36$ with the cubic form (first-step secant $1.32$), and again $0.52$ and $0.72$ on the approach to $\Qv$ [plus signs in Fig.~\ref{fig:hydro}(b)]. These bounds agree with the SAC peak positions to two digits, and on the $XY$ sound branch [$1.12$, $2.07$ and $2.82$ at the same three axial momenta and $1.57$ at $(\tfrac\pi8,\tfrac\pi8)$] to within $0.04$, so the soft branch is in the data and not a continuation artifact. On the plateau, where no sharp mode exists and the bound lies inside a continuum, the two differ by up to $0.41$. An upper bound cannot by itself prove the IS2 gap, but the large values (nothing below $2.59$), the SAC edges, and equal-time weights an order of magnitude smaller than in the supersolid together leave no room for a soft branch.

\paragraph{Rotor tower.} The tower is not frozen out at $\beta=32$, and that is visible in the raw data: $\Gxx(\Qv,\tau)/\Gxx(\Qv,0)$ falls to $0.873$ (SS) and $0.729$ ($XY$) at the largest $\tau$, $15.6\simeq\beta/2$, rather than staying constant. The rotor model predicts this decay with no free parameter. Thermally averaging over the sectors $M=S^z_{\rm tot}$ with $E(M)=\Drot(M-Nm_z)^2$ and Boltzmann weights $p_M=e^{-\beta E(M)}$, and letting $S^\pm$ connect $M\to M\pm1$ with an $M$-independent matrix element,
\begin{equation}
  \frac{\Gxx_{\rm rot}(\tau)}{\Gxx_{\rm rot}(0)}
  =\frac{\sum_M p_M\bigl[e^{-\tau\varepsilon^+_M}+e^{-\tau\varepsilon^-_M}\bigr]}{2\sum_M p_M},
  \label{eq:rotorG}
\end{equation}
with $\varepsilon^\pm_M=E(M\pm1)-E(M)$, gives $0.891$ (SS) and $0.736$ ($XY$) there and follows the measured curves over the whole window to within $0.02$ at $\Qv$ and $0.04$ at $(0,0)$ in the supersolid and $0.01$ in the $XY$ phase [Fig.~\ref{fig:hydro}(a)]. The small deficit is set already at $\tau\lesssim0.5$ and is $\tau$ independent thereafter: it is the $2\%$ ($\Qv$, SS), $4\%$ [$(0,0)$, SS] and $1\%$ ($XY$) of $\Sxx(\qv)$ that sits in the genuinely inelastic spectrum. In the supersolid this share includes a weak maximum at $\omega=1.19$ at $\Qv$ [Fig.~\ref{fig:cuts}(c)] and at $0.90$ at $(0,0)$, each with $0.3\%$ of $\Sxx(\qv)$, to which we attach no interpretation. The same model accounts for a second number. Because the tower is thermally populated, part of its weight sits at $\omega<0$, and by Eq.~\eqref{eq:sumrule} the positive-frequency weight falls below $\pi\Sxx(\qv)$ by exactly that share: the rotor sum, evaluated with $m_z=0.3597$ and $0.7063$ (the fraction depends on $Nm_z$ modulo 1, so four digits are needed), predicts positive-frequency fractions of $0.691$ (SS) and $0.799$ ($XY$), against $0.697$ at $\Qv$ and $0.707$ at $(0,0)$ in the supersolid and $0.786$ in the $XY$ phase, all within $0.02$, while at every other momentum the ratio is $1.0000$. In the supersolid the normalized $\Gxx(\qv,\tau)$ at $\qv=(0,0)$ has the same $\tau$ dependence as at $\Qv$ (local log slopes identical to three digits), as folding requires. That the supersolid carries both a doublet in $\Szz$ and a tower in $\Sxx$ at the same $\Qv$ is the finite-size statement of supersolidity, and the $\chizz$ that sets the rotor gap is the same one that fixes the Goldstone velocity and the sound-mode weight.

\section{Summary and discussion}
\label{sec:discussion}

We have computed the dynamic structure factors of the square-lattice $S=1$ antiferromagnet for the model of Fig.~1 of the Letter, one field each in the supersolid, the IS2 plateau and the canted $XY$ phase. The supersolid has one Goldstone mode that the checkerboard folds onto both $\Qv$ and $(0,0)$; it appears in both channels with a single velocity, $v=1.36\pm0.01$, equal within $1\%$ to $\sqrt{\rhos/\chizz}$ from the same runs. The $XY$ phase shows the mode only at $\Qv$ in $\Sxx$ and at $(0,0)$ in $\Szz$, and the plateau shows a gapped magnon band and a two-magnon longitudinal continuum. Exactly at the special wave vectors the finite lattice shows a $1/N$ rotor tower in $\Sxx$ and an exponentially split Ising doublet in $\Szz$, the supersolid carrying both at $\Qv$. A fully gapped $\Szz$ is the signature of the incompressible plateau, not of the $Z_2$ order.

The two-peak transverse structure factor of the Letter records that the condensate is modulated by the checkerboard. The dynamics records the same folding and adds two constraints that a static peak does not carry: the branches at $\Qv$ and $(0,0)$, in both channels, share one velocity, and that velocity is fixed by $\rhos$ and $\chizz$. With the continuation-free decay rates, these comparisons are the strongest argument we can give that the soft branches are physical rather than artifacts of the continuation.

The single Goldstone mode is what distinguishes a lattice spin supersolid with a discrete density wave from the continuum supersolids of quantum gases, where broken translations supply a second gapless branch~\cite{Leonard2017,Tanzi2019,Guo2019,Natale2019}. On the square lattice the $Z_2$ order commutes with the $U(1)$ generator and contributes only a gapped amplitude sector, visible here as the $3.9$ edge of the connected $\Szz(\Qv,\omega)$. The triangular-lattice cobaltates studied experimentally show broad continua in their zero-field supersolid phases~\cite{Zhu2024,Sheng2025}, attributed to magnon decay on the frustrated lattice~\cite{Sheng2025,Zhu2024} and possibly to proximity to a Dirac spin liquid~\cite{Zhu2024}. Quantum Monte Carlo with stochastic analytic continuation resolves continua of that kind where they occur, in one dimension~\cite{Yang2025,Yang2026} (there with parametrizations adapted to an edge or a quasiparticle pole) and at a two-dimensional deconfined quantum critical point~\cite{Yang2025b}, so their absence at low energy in the unfrustrated $S=1$ model is a property of the model and not of the method. The supersolid is nevertheless the least coherent of the three phases in the transverse channel, with about $78\%$ of the weight in continua at $4.5$--$14$ and only $21$--$24\%$ in the branch on its flat top, against $32$--$49\%$ in the $XY$ phase away from $\Qv$ and $(0,0)$ and $48$--$92\%$ on the plateau over the whole path.

For experiments, the result suggests where to look. In a material described by Eq.~\eqref{eq:H}, inelastic neutron scattering in the supersolid field window should find a linearly dispersing mode emerging from both the antiferromagnetic and the nuclear zone center, with the same velocity in the two channels, and the measured differential susceptibility $\chizz=dm_z/dB$ then converts that velocity into the spin stiffness through $\rhos=\chizz v^2$. The magnetic cross section carries the factor $\delta_{ab}-\hat q_a\hat q_b$, so $\Szz$ is invisible for scattering vectors along the field and the longitudinal mode must be sought with $\qv$ perpendicular or nearly perpendicular to it, and the identification of the two channels with spin-flip and non-spin-flip scattering holds only for a stated neutron polarization. On the plateau the same experiment should find a gapped band and a two-magnon continuum in the longitudinal channel; in the canted phase, a single Goldstone cone at the antiferromagnetic wave vector only.

\begin{acknowledgments}
This work was supported by the Singapore Ministry of Education (MOE) Academic Research Fund
Tier~3 Grant No.~MOE-MOET32023-0003 and Tier~2 Grant No.~MOE-T2EP50223-0019.
We would like to acknowledge the NTU High Performance Computing Centre (HPCC), Singapore, for providing the computing resources, facilities, and services that have contributed significantly to this work. We also acknowledge the use of computing resources at the National Supercomputing Centre (NSCC), Singapore.
\end{acknowledgments}

\section*{Data availability}

The data that support the findings of this article are not publicly available upon publication because it is not technically feasible and/or the cost of preparing, depositing, and hosting the data would be prohibitive within the terms of this research project. The data are available from the authors upon reasonable request.

\section*{AI use disclosure}

The research was conceived and directed by the authors; the choice of model and parameters, the quantum
Monte Carlo and stochastic analytic continuation calculations, and the physical interpretation of the
results are their own. Claude Code (Anthropic), with the Claude Opus~5, Claude Opus~5.5, Claude Fable~5.1 and Claude Fable~5.2 models, was used as
an assistive tool under the authors' direction: to help draft and edit the text from the authors' research
notes, to write the scripts that plot the figures, and to carry out consistency checks and auxiliary
analyses of the reported quantities, such as the parameter-free comparisons of Sec.~\ref{sec:checks}.
The authors reviewed each step, verified every number, statement and figure against the simulation data,
and take full responsibility for the content of this paper.

\appendix

\section{Validation and field scans}
\label{app:scan}

\begin{figure}[!t]
\centering
\includegraphics[width=\columnwidth]{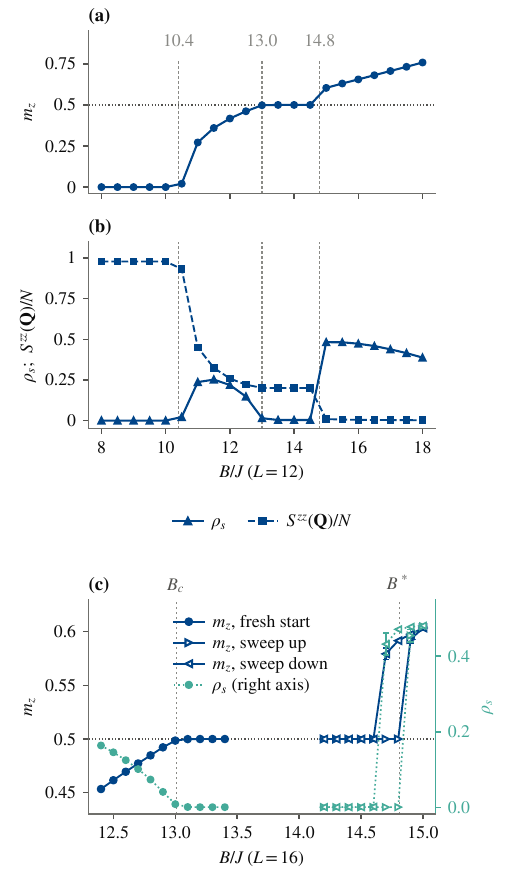}
\caption{Static field scans that locate the phase boundaries and the plateau edges. (a), (b) $L=12$, $\beta=32$, $D=3.0$, $J_z=3.6$, steps of $\Delta B=0.5$, with fields in the units of Eq.~\eqref{eq:H} in all panels. (a) Magnetization; the dotted horizontal line is $m_z=1/2$. (b) Spin stiffness (solid, triangles) and staggered structure factor (dashed, squares). Vertical dotted lines mark the transition fields $10.4$, $13.0$ and $14.8$ of Fig.~\ref{fig:phases}. (c) $L=16$, $\beta=32$, $D=3.0$, $J_z=3.6$, steps of $\Delta B=0.1$ across the two plateau edges: $m_z$ (left axis) and $\rhos$ (right axis, dotted). Across the continuous SS--IS2 edge every field is a fresh start; across the first-order IS2--$XY$ edge the field is swept up from $14.2$ and down from $15.0$, each point continuing from the previous one. Vertical lines: $B_c=13.01$ and the energy crossing $B^*=14.81$. Error bars in (c) are one standard error; they exceed the symbols only at the two switching points of the sweeps, $B=14.9$ (up) and $14.7$ (down). Statistical errors in (a) and (b) are smaller than the symbols.}
\label{fig:scan}
\end{figure}

Three checks validate our simulations and their energy unit. The first is absolute: the estimators used here were validated against dense exact diagonalization at this parameter point, entry by entry in imaginary time and distance and with no fitted constant ($2\times2$ clusters at $\beta=8$ and $32$, and a $4\times2$ cluster at $\beta=8$; root-mean-square normalized deviation $\simeq1$ over 328 entries), so our energy unit is the $J$ of Eq.~\eqref{eq:H} as written. The second is the phase diagram, Figs.~\ref{fig:scan}(a) and \ref{fig:scan}(b). The complete sequence of field-driven phases reported in the Letter is reproduced with increasing field strength: the Ising solid up to $B=10.0$, a supersolid window in which $m_z$, $\rhos$ and $\Szz(\Qv)/N$ are simultaneously finite, an exact $m_z=1/2$ plateau at $B=13.5$, $14.0$ and $14.5$, and a canted $XY$ state from $B=15.0$ onwards. The scan brackets the lower edge of the supersolid between $10.0$ and $10.5$; that this transition is of second order is taken from the Letter, the step of $0.5$ being too coarse to show it ($m_z$ rises from $0.02$ at $B=10.5$ to $0.27$ at $11.0$). The phase boundaries agree with those of Fig.~1 of the Letter, $5.2$, $6.5$ and $7.3$ on its field axis and twice that in the units of Eq.~\eqref{eq:H} (Sec.~\ref{sec:params}), to within the step $\Delta B=0.5$ of the scan. The third is the spin stiffness: converted to the same units, the $\rhos$ read from Fig.~1(d) of the Letter at the same magnetization agrees with ours to within the reading uncertainty ($\pm0.03$ on the axis of that figure, $\pm0.06$ in our units; Table~\ref{tab:static}), and the dimensionless quantities agree directly.

Figure~\ref{fig:scan}(c) refines the two plateau edges at $L=16$ in steps of $0.1$, with static observables only. Across the continuous SS--IS2 edge, $\tfrac12-m_z$ and $\rhos$ fall linearly, from $0.0152$ and $0.074$ at $B=12.8$ through $0.0077$ and $0.041$ at $12.9$ to $0.0014$ and $0.009$ at $13.0$; extrapolating the two linearly from $12.8$ and $12.9$ gives $B_c=13.00$ and $13.02$, i.e., $B_c=13.01\pm0.01$, the uncertainty being half the difference of the two extrapolations. The point at $13.0$ is not used, because the lattice then holds less than one hole on average [$N(\tfrac12-m_z)=3.9$, $2.0$ and $0.4$ at the three fields]. Across the first-order IS2--$XY$ edge the up and down sweeps show hysteresis: the plateau survives up to $14.8$ and the canted state down to $14.7$. The total energies of the two branches are linear in $B$ with slopes $-0.4998$ and $-0.599$, equal to $-m_z$ as they must be, and cross at $B^*=14.81\pm0.01$, where the uncertainty allows for the small entropy difference neglected at $T=1/32$. On the field axis of Fig.~1 of the Letter, $B^*$ is $7.4$ against the $7.3$ read from that figure. The difference lies within the hysteresis: in our downward sweep the plateau returns at $B=14.6$, which is $7.3$ on that axis.

\section{Analytic continuation: windows, normalization, and peak extraction}
\label{app:sac}

The number $N_\tau$ of imaginary-time points retained (relative error below $20\%$) varies strongly between data sets: its median over the momentum path is $41$ and $42$ for $\Sxx$ at $B=11.5$ and $14$ and $52$ for $\Szz$ at $B=11.5$, but only $19$ for $\Sxx$ at $B=17$, $21$ for $\Szz$ at $B=14$ and $30$ for $\Szz$ at $B=17$, with an overall range of $15$ to $61$; the short windows are those in which the correlator decays fastest or carries least weight. The median $\chi^2_{\min}/N_\tau$ is $0.76$--$0.94$ per data set.

Reconstructing $G$ from each spectrum through Eq.~\eqref{eq:norm} on its native $\Delta\omega=0.005$ grid reproduces the measured $G(\qv,0)$ to better than $3\times10^{-4}$ (to $10^{-9}$ away from the rotor-tower columns of Sec.~\ref{sec:sectors}) and $G(\qv,\tau)/G(\qv,0)$ to better than $5\times10^{-4}$ at every point $0.1\le\tau\le15.6$, for all 159 continued spectra of Fig.~\ref{fig:maps} (27 distinct momenta, six data sets, less the three elastic $\Szz$ columns at $(0,0)$). Bin-averaging a spectrum to $\Delta\omega=0.05$ before back-transforming would spoil $G(\qv,0)$ at the rotor-tower columns by $2$--$3\times10^{-2}$, because $e^{-(\beta-\tau)\omega}$ varies by a factor of five across one such bin at $\omega\simeq0.03$, so all spectra are used and plotted unbinned. The Bragg subtraction of Sec.~\ref{sec:sectors} also removes any inelastic weight at $\Qv$ lying below the resolution; such weight, if it lay above $\omega\simeq1/\beta$, would make $\Gzz(\Qv,\tau)$ vary between $\tau=2$ and $15.6$ by at least $8\%$ of itself, so the observed flatness bounds it to about a part in $10^6$ of the column, below $0.1\%$ of the connected weight.

All peak positions, weight fractions and edges quoted in the text come from one procedure with fixed thresholds. In the spectrum smoothed with a Gaussian of width $\sigma_\omega=0.05$, maxima with a prominence of at least $2\%$ of the maximum count as peaks. Because this cut scales with the tallest peak, a weak continuum next to a growing branch drops below it, as next to $\Qv$ in the supersolid; the window of each peak is therefore bounded by the minima of a finer peak set (prominence $0.2\%$), which keeps the weight of such a continuum out of the branch. Only the positions of these maxima and minima are taken from the smoothed spectrum; weights and edges use the unsmoothed one, so the broadening by the smoothing enters no quoted number. Weight outside the windows of the $2\%$ peaks is an unresolved remainder, so the fractions at one momentum need not sum to one. Neither choice is critical: for $0.01\le\sigma_\omega\le0.05$ the mean number of peaks per spectrum changes only from $1.74$ to $1.80$ and the lowest-peak positions move by at most $0.05$, and every lowest peak used in the text has prominence above $5\%$. An independent 1152-bin data set at $B=11.5$ gives the same peak positions; the least stable quantity in this comparison, the edge of the connected $\Szz(\Qv)$, is $4.2$ there against $3.9$ in the 4096-bin data.

\bibliography{refs}

@article{Sengupta2007,
  author  = {Sengupta, P. and Batista, C. D.},
  title   = {Field-Induced Supersolid Phase in Spin-One {H}eisenberg Models},
  journal = {Phys. Rev. Lett.},
  volume  = {98},
  pages   = {227201},
  year    = {2007},
  doi     = {10.1103/PhysRevLett.98.227201}
}

@article{Andreev1969,
  author  = {Andreev, A. F. and Lifshitz, I. M.},
  title   = {Quantum Theory of Defects in Crystals},
  journal = {Sov. Phys. JETP},
  volume  = {29},
  pages   = {1107},
  year    = {1969}
}

@article{Chester1970,
  author  = {Chester, G. V.},
  title   = {Speculations on {B}ose-{E}instein Condensation and Quantum Crystals},
  journal = {Phys. Rev. A},
  volume  = {2},
  pages   = {256},
  year    = {1970},
  doi     = {10.1103/PhysRevA.2.256}
}

@article{Leggett1970,
  author  = {Leggett, A. J.},
  title   = {Can a Solid Be ``Superfluid''?},
  journal = {Phys. Rev. Lett.},
  volume  = {25},
  pages   = {1543},
  year    = {1970},
  doi     = {10.1103/PhysRevLett.25.1543}
}

@article{Boninsegni2012,
  author  = {Boninsegni, Massimo and Prokof'ev, Nikolay V.},
  title   = {Colloquium: Supersolids: What and where are they?},
  journal = {Rev. Mod. Phys.},
  volume  = {84},
  pages   = {759},
  year    = {2012},
  doi     = {10.1103/RevModPhys.84.759}
}

@article{Kim2004,
  author  = {Kim, E. and Chan, M. H. W.},
  title   = {Probable observation of a supersolid helium phase},
  journal = {Nature},
  volume  = {427},
  pages   = {225},
  year    = {2004},
  doi     = {10.1038/nature02220}
}

@article{Leonard2017,
  author  = {L{\'e}onard, Julian and Morales, Andrea and Zupancic, Philip and Donner, Tobias and Esslinger, Tilman},
  title   = {Monitoring and manipulating {H}iggs and {G}oldstone modes in a supersolid quantum gas},
  journal = {Science},
  volume  = {358},
  pages   = {1415},
  year    = {2017},
  doi     = {10.1126/science.aan2608}
}

@article{Tanzi2019,
  author  = {Tanzi, L. and Roccuzzo, S. M. and Lucioni, E. and Fam{\`a}, F. and Fioretti, A. and Gabbanini, C. and Modugno, G. and Recati, A. and Stringari, S.},
  title   = {Supersolid symmetry breaking from compressional oscillations in a dipolar quantum gas},
  journal = {Nature},
  volume  = {574},
  pages   = {382},
  year    = {2019},
  doi     = {10.1038/s41586-019-1568-6}
}

@article{Guo2019,
  author  = {Guo, Mingyang and B{\"o}ttcher, Fabian and Hertkorn, Jens and Schmidt, Jan-Niklas and Wenzel, Matthias and B{\"u}chler, Hans Peter and Langen, Tim and Pfau, Tilman},
  title   = {The low-energy {G}oldstone mode in a trapped dipolar supersolid},
  journal = {Nature},
  volume  = {574},
  pages   = {386},
  year    = {2019},
  doi     = {10.1038/s41586-019-1569-5}
}

@article{Natale2019,
  author  = {Natale, G. and van Bijnen, R. M. W. and Patscheider, A. and Petter, D. and Mark, M. J. and Chomaz, L. and Ferlaino, F.},
  title   = {Excitation Spectrum of a Trapped Dipolar Supersolid and Its Experimental Evidence},
  journal = {Phys. Rev. Lett.},
  volume  = {123},
  pages   = {050402},
  year    = {2019},
  doi     = {10.1103/PhysRevLett.123.050402}
}

@article{Matsubara1956,
  author  = {Matsubara, Takeo and Matsuda, Hirotsugu},
  title   = {A Lattice Model of Liquid Helium, {I}},
  journal = {Prog. Theor. Phys.},
  volume  = {16},
  pages   = {569},
  year    = {1956},
  doi     = {10.1143/PTP.16.569}
}

@article{Giamarchi2008,
  author  = {Giamarchi, Thierry and R{\"u}egg, Christian and Tchernyshyov, Oleg},
  title   = {{B}ose-{E}instein condensation in magnetic insulators},
  journal = {Nat. Phys.},
  volume  = {4},
  pages   = {198},
  year    = {2008},
  doi     = {10.1038/nphys893}
}

@article{Zapf2014,
  author  = {Zapf, Vivien and Jaime, Marcelo and Batista, C. D.},
  title   = {Bose-{E}instein condensation in quantum magnets},
  journal = {Rev. Mod. Phys.},
  volume  = {86},
  pages   = {563},
  year    = {2014},
  doi     = {10.1103/RevModPhys.86.563}
}

@article{Wessel2005,
  author  = {Wessel, Stefan and Troyer, Matthias},
  title   = {Supersolid Hard-Core Bosons on the Triangular Lattice},
  journal = {Phys. Rev. Lett.},
  volume  = {95},
  pages   = {127205},
  year    = {2005},
  doi     = {10.1103/PhysRevLett.95.127205}
}

@article{Heidarian2005,
  author  = {Heidarian, Dariush and Damle, Kedar},
  title   = {Persistent Supersolid Phase of Hard-Core Bosons on the Triangular Lattice},
  journal = {Phys. Rev. Lett.},
  volume  = {95},
  pages   = {127206},
  year    = {2005},
  doi     = {10.1103/PhysRevLett.95.127206}
}

@article{Melko2005,
  author  = {Melko, R. G. and Paramekanti, A. and Burkov, A. A. and Vishwanath, A. and Sheng, D. N. and Balents, L.},
  title   = {Supersolid Order from Disorder: Hard-Core Bosons on the Triangular Lattice},
  journal = {Phys. Rev. Lett.},
  volume  = {95},
  pages   = {127207},
  year    = {2005},
  doi     = {10.1103/PhysRevLett.95.127207}
}

@article{Boninsegni2005,
  author  = {Boninsegni, Massimo and Prokof'ev, Nikolay},
  title   = {Supersolid Phase of Hard-Core Bosons on a Triangular Lattice},
  journal = {Phys. Rev. Lett.},
  volume  = {95},
  pages   = {237204},
  year    = {2005},
  doi     = {10.1103/PhysRevLett.95.237204}
}

@article{Batrouni2000,
  author  = {Batrouni, G. G. and Scalettar, R. T.},
  title   = {Phase Separation in Supersolids},
  journal = {Phys. Rev. Lett.},
  volume  = {84},
  pages   = {1599},
  year    = {2000},
  doi     = {10.1103/PhysRevLett.84.1599}
}

@article{Sengupta2005,
  author  = {Sengupta, Pinaki and Pryadko, Leonid P. and Alet, Fabien and Troyer, Matthias and Schmid, Guido},
  title   = {Supersolids versus Phase Separation in Two-Dimensional Lattice Bosons},
  journal = {Phys. Rev. Lett.},
  volume  = {94},
  pages   = {207202},
  year    = {2005},
  doi     = {10.1103/PhysRevLett.94.207202}
}

@article{Ng2006,
  author  = {Ng, Kwai-Kong and Lee, T. K.},
  title   = {Supersolid Phase in Spin Dimer {XXZ} Systems under a Magnetic Field},
  journal = {Phys. Rev. Lett.},
  volume  = {97},
  pages   = {127204},
  year    = {2006},
  doi     = {10.1103/PhysRevLett.97.127204}
}

@article{Laflorencie2007,
  author  = {Laflorencie, Nicolas and Mila, Fr{\'e}d{\'e}ric},
  title   = {Quantum and Thermal Transitions Out of the Supersolid Phase of a {2D} Quantum Antiferromagnet},
  journal = {Phys. Rev. Lett.},
  volume  = {99},
  pages   = {027202},
  year    = {2007},
  doi     = {10.1103/PhysRevLett.99.027202}
}

@article{Sengupta2007b,
  author  = {Sengupta, P. and Batista, C. D.},
  title   = {Spin Supersolid in an Anisotropic Spin-One {H}eisenberg Chain},
  journal = {Phys. Rev. Lett.},
  volume  = {99},
  pages   = {217205},
  year    = {2007},
  doi     = {10.1103/PhysRevLett.99.217205}
}

@article{Peters2009,
  author  = {Peters, D. and McCulloch, I. P. and Selke, W.},
  title   = {Spin-one {H}eisenberg antiferromagnetic chain with exchange and single-ion anisotropies},
  journal = {Phys. Rev. B},
  volume  = {79},
  pages   = {132406},
  year    = {2009},
  doi     = {10.1103/PhysRevB.79.132406}
}

@article{Rossini2011,
  author  = {Rossini, Davide and Giovannetti, Vittorio and Fazio, Rosario},
  title   = {Spin-supersolid phase in {H}eisenberg chains: A characterization via matrix product states with periodic boundary conditions},
  journal = {Phys. Rev. B},
  volume  = {83},
  pages   = {140411},
  year    = {2011},
  doi     = {10.1103/PhysRevB.83.140411}
}

@article{Yamamoto2014,
  author  = {Yamamoto, Daisuke and Marmorini, Giacomo and Danshita, Ippei},
  title   = {Quantum Phase Diagram of the Triangular-Lattice {XXZ} Model in a Magnetic Field},
  journal = {Phys. Rev. Lett.},
  volume  = {112},
  pages   = {127203},
  year    = {2014},
  doi     = {10.1103/PhysRevLett.112.127203}
}

@article{Gao2022,
  author  = {Gao, Yuan and Fan, Yu-Chen and Li, Han and Yang, Fan and Zeng, Xu-Tao and Sheng, Xian-Lei and Zhong, Ruidan and Qi, Yang and Wan, Yuan and Li, Wei},
  title   = {Spin supersolidity in nearly ideal easy-axis triangular quantum antiferromagnet {N}a$_2${B}a{C}o({PO}$_4$)$_2$},
  journal = {npj Quantum Mater.},
  volume  = {7},
  pages   = {89},
  year    = {2022},
  doi     = {10.1038/s41535-022-00500-3}
}

@article{Xiang2024,
  author  = {Xiang, Junsen and Zhang, Chuandi and Gao, Yuan and Schmidt, Wolfgang and Schmalzl, Karin and Wang, Chin-Wei and Li, Bo and Xi, Ning and Liu, Xin-Yang and Jin, Hai and Li, Gang and Shen, Jun and Chen, Ziyu and Qi, Yang and Wan, Yuan and Jin, Wentao and Li, Wei and Sun, Peijie and Su, Gang},
  title   = {Giant magnetocaloric effect in spin supersolid candidate {N}a$_2${B}a{C}o({PO}$_4$)$_2$},
  journal = {Nature},
  volume  = {625},
  pages   = {270},
  year    = {2024},
  doi     = {10.1038/s41586-023-06885-w}
}

@article{Sheng2025,
  author  = {Sheng, Jieming and Wang, Le and Jiang, Wenrui and Ge, Han and Zhao, Nan and Li, Tiantian and Kofu, Maiko and Yu, Dehong and Zhu, Wei and Mei, Jia-Wei and Wang, Zhentao and Wu, Liusuo},
  title   = {Continuum of spin excitations in an ordered magnet},
  journal = {The Innovation},
  volume  = {6},
  pages   = {100769},
  year    = {2025},
  doi     = {10.1016/j.xinn.2024.100769}
}

@article{Zhu2024,
  author  = {Zhu, M. and Romerio, V. and Steiger, N. and Nabi, S. D. and Murai, N. and Ohira-Kawamura, S. and Povarov, K. Yu. and Skourski, Y. and Sibille, R. and Keller, L. and Yan, Z. and Gvasaliya, S. and Zheludev, A.},
  title   = {Continuum Excitations in a Spin Supersolid on a Triangular Lattice},
  journal = {Phys. Rev. Lett.},
  volume  = {133},
  pages   = {186704},
  year    = {2024},
  doi     = {10.1103/PhysRevLett.133.186704}
}

@article{Sandvik1991,
  author  = {Sandvik, Anders W. and Kurkij{\"a}rvi, Juhani},
  title   = {Quantum {M}onte {C}arlo simulation method for spin systems},
  journal = {Phys. Rev. B},
  volume  = {43},
  pages   = {5950},
  year    = {1991},
  doi     = {10.1103/PhysRevB.43.5950}
}

@article{Sandvik1999,
  author  = {Sandvik, Anders W.},
  title   = {Stochastic series expansion method with operator-loop update},
  journal = {Phys. Rev. B},
  volume  = {59},
  pages   = {R14157},
  year    = {1999},
  doi     = {10.1103/PhysRevB.59.R14157}
}

@article{Syljuasen2002,
  author  = {Sylju{\aa}sen, Olav F. and Sandvik, Anders W.},
  title   = {Quantum {M}onte {C}arlo with directed loops},
  journal = {Phys. Rev. E},
  volume  = {66},
  pages   = {046701},
  year    = {2002},
  doi     = {10.1103/PhysRevE.66.046701}
}

@article{Dorneich2001,
  author  = {Dorneich, Ansgar and Troyer, Matthias},
  title   = {Accessing the dynamics of large many-particle systems using the stochastic series expansion},
  journal = {Phys. Rev. E},
  volume  = {64},
  pages   = {066701},
  year    = {2001},
  doi     = {10.1103/PhysRevE.64.066701}
}

@article{Sandvik1998,
  author  = {Sandvik, Anders W.},
  title   = {Stochastic method for analytic continuation of quantum {M}onte {C}arlo data},
  journal = {Phys. Rev. B},
  volume  = {57},
  pages   = {10287},
  year    = {1998},
  doi     = {10.1103/PhysRevB.57.10287}
}

@misc{Beach2004,
  author        = {Beach, K. S. D.},
  title         = {Identifying the maximum entropy method as a special limit of stochastic analytic continuation},
  year          = {2004},
  eprint        = {cond-mat/0403055},
  archivePrefix = {arXiv}
}

@article{Sandvik2016,
  author  = {Sandvik, Anders W.},
  title   = {Constrained sampling method for analytic continuation},
  journal = {Phys. Rev. E},
  volume  = {94},
  pages   = {063308},
  year    = {2016},
  doi     = {10.1103/PhysRevE.94.063308}
}

@article{Shao2017,
  author  = {Shao, Hui and Qin, Yan Qi and Capponi, Sylvain and Chesi, Stefano and Meng, Zi Yang and Sandvik, Anders W.},
  title   = {Nearly Deconfined Spinon Excitations in the Square-Lattice Spin-1/2 {H}eisenberg Antiferromagnet},
  journal = {Phys. Rev. X},
  volume  = {7},
  pages   = {041072},
  year    = {2017},
  doi     = {10.1103/PhysRevX.7.041072}
}

@article{Shao2023,
  author  = {Shao, Hui and Sandvik, Anders W.},
  title   = {Progress on stochastic analytic continuation of quantum {M}onte {C}arlo data},
  journal = {Phys. Rep.},
  volume  = {1003},
  pages   = {1},
  year    = {2023},
  doi     = {10.1016/j.physrep.2022.11.002}
}

@article{Jarrell1996,
  author  = {Jarrell, Mark and Gubernatis, J. E.},
  title   = {Bayesian inference and the analytic continuation of imaginary-time quantum {M}onte {C}arlo data},
  journal = {Phys. Rep.},
  volume  = {269},
  pages   = {133},
  year    = {1996},
  doi     = {10.1016/0370-1573(95)00074-7}
}

@article{Anderson1952,
  author  = {Anderson, P. W.},
  title   = {An Approximate Quantum Theory of the Antiferromagnetic Ground State},
  journal = {Phys. Rev.},
  volume  = {86},
  pages   = {694},
  year    = {1952},
  doi     = {10.1103/PhysRev.86.694}
}

@article{Bernu1992,
  author  = {Bernu, B. and Lhuillier, C. and Pierre, L.},
  title   = {Signature of {N}\'eel order in exact spectra of quantum antiferromagnets on finite lattices},
  journal = {Phys. Rev. Lett.},
  volume  = {69},
  pages   = {2590},
  year    = {1992},
  doi     = {10.1103/PhysRevLett.69.2590}
}

@misc{Lhuillier2005,
  author        = {Lhuillier, Claire},
  title         = {Frustrated Quantum Magnets},
  year          = {2005},
  eprint        = {cond-mat/0502464},
  archivePrefix = {arXiv}
}

@article{Tasaki2019,
  author  = {Tasaki, Hal},
  title   = {Long-Range Order, ``Tower'' of States, and Symmetry Breaking in Lattice Quantum Systems},
  journal = {J. Stat. Phys.},
  volume  = {174},
  pages   = {735},
  year    = {2019},
  doi     = {10.1007/s10955-018-2193-8}
}

@article{Halperin1969,
  author  = {Halperin, B. I. and Hohenberg, P. C.},
  title   = {Hydrodynamic Theory of Spin Waves},
  journal = {Phys. Rev.},
  volume  = {188},
  pages   = {898},
  year    = {1969},
  doi     = {10.1103/PhysRev.188.898}
}

@article{Goldstone1961,
  author  = {Goldstone, J.},
  title   = {Field theories with ``Superconductor'' solutions},
  journal = {Il Nuovo Cimento},
  volume  = {19},
  pages   = {154},
  year    = {1961},
  doi     = {10.1007/BF02812722}
}

@article{Watanabe2012,
  author  = {Watanabe, Haruki and Murayama, Hitoshi},
  title   = {Unified Description of {N}ambu-{G}oldstone Bosons without {L}orentz Invariance},
  journal = {Phys. Rev. Lett.},
  volume  = {108},
  pages   = {251602},
  year    = {2012},
  doi     = {10.1103/PhysRevLett.108.251602}
}

@article{Penrose1956,
  author  = {Penrose, Oliver and Onsager, Lars},
  title   = {Bose-{E}instein Condensation and Liquid Helium},
  journal = {Phys. Rev.},
  volume  = {104},
  pages   = {576},
  year    = {1956},
  doi     = {10.1103/PhysRev.104.576}
}

@article{Zhitomirsky1998,
  author  = {Zhitomirsky, M. E. and Nikuni, T.},
  title   = {Magnetization curve of a square-lattice {H}eisenberg antiferromagnet},
  journal = {Phys. Rev. B},
  volume  = {57},
  pages   = {5013},
  year    = {1998},
  doi     = {10.1103/PhysRevB.57.5013}
}

@article{Zhitomirsky2013,
  author  = {Zhitomirsky, M. E. and Chernyshev, A. L.},
  title   = {Colloquium: Spontaneous magnon decays},
  journal = {Rev. Mod. Phys.},
  volume  = {85},
  pages   = {219},
  year    = {2013},
  doi     = {10.1103/RevModPhys.85.219}
}

@misc{Flynn2026,
  author        = {Flynn, Ryan and Iaizzi, Adam and Yang, Sibin and Kao, Ying-Jer and Sandvik, Anders W.},
  title         = {Imaginary-time correlations in time-sliced stochastic series expansion},
  year          = {2026},
  eprint        = {2608.13477},
  archivePrefix = {arXiv},
  primaryClass  = {cond-mat.str-el}
}

@article{Schumm2024,
  author  = {Schumm, Gabe and Yang, Sibin and Sandvik, Anders W.},
  title   = {Cross validation in stochastic analytic continuation},
  journal = {Phys. Rev. E},
  volume  = {110},
  pages   = {055307},
  year    = {2024},
  doi     = {10.1103/PhysRevE.110.055307}
}

@article{Yang2025,
  author  = {Yang, Sibin and Schumm, Gabe and Sandvik, Anders W.},
  title   = {Dynamic structure factor of a spin-1/2 {H}eisenberg chain with long-range interactions},
  journal = {Phys. Rev. B},
  volume  = {111},
  pages   = {224404},
  year    = {2025},
  doi     = {10.1103/PhysRevB.111.224404}
}

@article{Yang2026,
  author  = {Yang, Sibin and Schumm, Gabe and Zhao, Bowen and Sandvik, Anders W.},
  title   = {Single-hole spectral functions in one-dimensional quantum magnets with different ground states},
  journal = {Phys. Rev. B},
  volume  = {114},
  pages   = {055132},
  year    = {2026},
  doi     = {10.1103/y3sq-8knz}
}

@misc{Yang2025b,
  author        = {Yang, Sibin and Sandvik, Anders W.},
  title         = {Spinons and spin-charge separation at the deconfined quantum critical point},
  year          = {2025},
  eprint        = {2512.02962},
  archivePrefix = {arXiv},
  primaryClass  = {cond-mat.str-el}
}

@phdthesis{YangThesis,
  author = {Yang, Sibin},
  title  = {Spectral properties of low-dimensional quantum spin systems},
  school = {Boston University},
  year   = {2026}
}

@article{Kosterlitz1973,
  author  = {Kosterlitz, J. M. and Thouless, D. J.},
  title   = {Ordering, metastability and phase transitions in two-dimensional systems},
  journal = {J. Phys. C: Solid State Phys.},
  volume  = {6},
  pages   = {1181},
  year    = {1973},
  doi     = {10.1088/0022-3719/6/7/010}
}

@article{Nelson1977,
  author  = {Nelson, David R. and Kosterlitz, J. M.},
  title   = {Universal Jump in the Superfluid Density of Two-Dimensional Superfluids},
  journal = {Phys. Rev. Lett.},
  volume  = {39},
  pages   = {1201},
  year    = {1977},
  doi     = {10.1103/PhysRevLett.39.1201}
}

@article{Murakami2013,
  author  = {Murakami, Yuta and Oka, Takashi and Aoki, Hideo},
  title   = {Supersolid states in a spin system: Phase diagram and collective excitations},
  journal = {Phys. Rev. B},
  volume  = {88},
  pages   = {224404},
  year    = {2013},
  doi     = {10.1103/PhysRevB.88.224404}
}

\end{document}